\documentclass[twocolumn,pra,amsmath,amssymb,showpacs,
superscriptaddress]{revtex4}
\usepackage{epsfig}
\usepackage[dvipsnames]{xcolor}
\usepackage{color}
\usepackage{graphicx}
\usepackage{dcolumn}
\usepackage{bm}
\usepackage{braket}
\usepackage{amsthm}
\usepackage[colorlinks=true,
           linkcolor=red,
            urlcolor=blue,
           citecolor=ForestGreen]{hyperref}
\usepackage{array}
\usepackage{booktabs, multirow}
\usepackage{diagbox}
\usepackage{pict2e}
\usepackage{comment}
\usepackage{ulem}
\providecommand{\openone}{\leavevmode\hbox{\small1\kern-3.8pt\normalsize1}}

\begin{document}
\title{Work Fluctuations in Bosonic Josephson Junctions}

\author{R. G. Lena}
\affiliation{Dipartimento di Fisica e Chimica, Universit\`{a} di Palermo, via Archirafi 36, 90123 Palermo, Italy}
\author{G. M. Palma}
\affiliation{NEST, Istituto Nanoscienze-CNR and Dipartimento di Fisica e Chimica, Universit\`{a} degli Studi di Palermo, via Archirafi 36, I-90123 Palermo, Italy}
\author{G. De Chiara}
\affiliation{Centre for Theoretical Atomic, Molecular and Optical Physics, School of Mathematics and Physics, Queen's University, Belfast BT7 1NN, United Kingdom}

\date{\today }

\begin{abstract}
We calculate the first two moments and full probability distribution of the work performed on a system of bosonic particles in a two-mode Bose-Hubbard Hamiltonian when the self-interaction term is varied instantaneously or with a finite-time ramp. In the instantaneous case, we show how the irreversible work scales differently depending on whether the system is driven to the Josephson or Fock regime of the bosonic Josephson junction. In the finite-time case, we use optimal control techniques to substantially decrease the irreversible work to negligible values. Our analysis can be implemented in present-day experiments with ultracold atoms and we show how to relate the work statistics to that of the population imbalance of the two modes.
\end{abstract}


\maketitle

\section{Introduction}
Thermodynamics has lasted through all the scientific revolutions that have occurred in the last centuries. 
In the early days, thermodynamics was applied to macroscopic systems with a number of particles of the order or larger than the Avogadro number. This implies that when repeating a thermodynamic process under the same conditions, the observed values of thermodynamic quantities such as work, entropy and heat would always be the same. 
Recently, motivated by experiments in mesoscopic systems in solid state physics, molecular biology and in optical and atomic physics, attention has been turned to the {\it fluctuations} of thermodynamic quantities satisfying fundamental theorems \cite{jarzynski1997nonequilibrium,Crooks}. Such fluctuations can have a two-fold origin: they can be merely due to the smallness of mesoscopic systems giving rise to classical statistical fluctuations; or, they can be intrinsically quantum fluctuations. 

In the context of work in quantum mechanics, it has been shown that work cannot be identified with a single observable but rather to a generalised measurement \cite{TalknerPRE2007, CampisiRMP,Watanabe,RoncagliaPRL,DeChiaraPaz,TalknerHanggi2015}. Quantum fluctuations of work, in contrast to classical thermal fluctuations, survive when the temperature is lowered close to absolute zero. Their origins can be traced to the non-commutativity of operators in quantum mechanics: they emerge when driving a system with a sequence of Hamiltonians that do not commute with each other \cite{Fusco}. Such observation leaves an open question: how can one access the quantum fluctuations of work for a quantum mesoscopic system?

The aim of this paper is to answer positively to this question by studying the fluctuations of work generated by or made on a system of ultracold atoms in a double well potential. Recent technological and experimental progress in the field of cold atomic gases has triggered  enormous research activity towards the realisation of quantum simulators of condensed-matter physics models, quantum metrology and quantum information processors \cite{BlochRMP, NoriRMP, lewenstein2012ultracold}. 
Far less attention has been devoted to applications of out-of-equilibrium thermodynamics in ultracold atoms \cite{DeChiaraPaz} with the exception of the issue of thermalisation in closed quantum systems \cite{eisert2015quantum,PolkoRMP}.

Here, we consider a zero-temperature bosonic gas subject to a double-well potential. In the so-called two-mode approximation, the system can be regarded as a bosonic Josephson junction \cite{josephson1962possible,javanainen1986oscillatory} and its physics has been extensively studied both theoretically \cite{Giovanazzi,Milburn,LeggettRMP,JuliaDiaz,Barzanjeh} and experimentally \cite{GatiPRL,GatiJPB,Shin,schumm2005matter,Zibold,Maussang}. We calculate the work fluctuations in such setup after changing the inter-particle interaction strength by means of a Feshbach resonance. Similar effects could be obtained by changing the potential, raising or lowering the barrier separating the two wells.
For slow adiabatic changes of the interaction, the work needed to drive the system is approximately given by the free energy difference $\Delta F$ of the initial and final equilibrium states. However for fast driving the average work is always larger than  $\Delta F$ and their difference gives the irreversible work. 
We analyse the dependence of the irreversible work on the initial and final values of the self-interaction constant spanning the Rabi, Josephson and Fock regime of the double-well system. 

Furthermore, with the aim of reducing the irreversible work production, we employ optimal control methods to find a tailored time-dependence of the self-interaction \cite{CRAB}. We find that the irreversible work can be effectively reduced to a negligible value even if driving the system at a finite speed, challenging the minimal work principle \cite{Allahverdyan2005}. We test the robustness of our protocol to imperfections in the values of the self-interaction.

Our results can be tested in present-day experiments with ultracold atoms in double-well potentials \cite{GatiPRL,GatiJPB,Shin,schumm2005matter,Zibold,Maussang} or realising instances of the Lipkin-Meshkov-Glick (LMG) \cite{lipkin1965validity,Vidal} as for instance Bose-Einstein condensates in optical cavities \cite{brennecke2013real}. In the conclusions, we discuss a scheme to estimate work fluctuations in such systems.

\section{Model}

The system considered is a zero-temperature Bose-Einstein condensate in a double-well potential. For our study we use the two-mode Bose-Hubbard Hamiltonian
\begin{equation}
\hat{H}=\frac{U}{2} \left[\hat{n}_L(\hat{n}_L-1)+\hat{n}_R(\hat{n}_R-1)\right] 
 -J(\hat{a}_L\hat{a}_R^{\dagger}+\hat{a}_R\hat{a}_L^{\dagger}),
\end{equation}
where $J$ and $U$ are respectively the tunneling and the self-interaction energies and the number of particles $N$ is assumed to be constant. The operators $\hat{a}_L$ and $\hat{a}_R$ are the particle annihilation operators in the left and right well, respectively, and $\hat{n}_L$  and $\hat{n}_R$ are the corresponding number operators.

We analyse the system by using both a numerical and an analytical approach in order to find the eigenstates of the Hamiltonian. The analytical results are obtained by mapping the double-well to a quantum harmonic oscillator (QHO). In order to do that, we introduce the Schwinger operators
\begin{align}
\hat{J}_x&=\frac{1}{2}(\hat{a}_R^{\dagger}\hat{a}_R-\hat{a}_L^{\dagger}\hat{a}_L)\\
\hat{J}_y&=\frac{i}{2}(\hat{a}_R^{\dagger}\hat{a}_L-\hat{a}_L^{\dagger}\hat{a}_R)\\
\hat{J}_z&=\frac{1}{2}(\hat{a}_L^{\dagger}\hat{a}_R+\hat{a}_R^{\dagger}\hat{a}_L)
\end{align}
fulfilling the standard angular momentum commutation relations: $[\hat J_x,\hat J_y]= i\hbar \hat J_z$. These operators allow us to describe the system with the angular momentum formalism obtaining the Hamiltonian in the form
\begin{equation}
\hat{H}=-\frac{U}{2}N+U\left(\frac{N}{2}\right)^2+U\hat{J}_x^2-2J\hat{J}_z.
\label{H_Schwinger}
\end{equation}
We now map Hamiltonian \eqref{H_Schwinger} into that of a QHO employing the Holstein-Primakoff approximation, valid in the Josephson and Rabi regimes \cite{GatiJPB}. 
We remind that the system is in the Rabi and the Josephson regimes respectively when the conditions $\frac{UN}{J}\ll 1$ and $1\ll\frac{UN}{J}\ll N^2$ are fulfilled, whereas for $\frac{UN}{J}\gg N^2$ the system is in the Fock regime.
The operator $\hat{J}_x$, proportional to the population imbalance, is related to the position operator of the QHO by the equation $\hat{J}_x=\sqrt{N/2}\;\hat{x}$, whereas for the momentum $\hat{p}$ of the QHO it holds the relation $\hat{J}_y=-\sqrt{N/2}\;\hat{p}$. Hence the double-well system can be mapped to the QHO Hamiltonian
\begin{equation}
\hat{H}=E'+\frac{1}{2}m\omega_p^2\hat{x}^2+\frac{1}{2m}\hat{p}^2
\label{QHO_Hamiltonian}
\end{equation}
having an effective mass $m=(2J)^{-1}$ and the ``plasma frequency'' $\omega_p=2J\sqrt{\frac{UN}{2J}+1}$, where we defined $E'=-U\frac{N}{2}+U\frac{N^2}{4}-J-JN$.
While the mapping to Hamiltonian \eqref{H_Schwinger} is exact, the mapping to the QHO is only approximate and valid as long as $\langle \hat a^\dagger \hat a \rangle \ll N $ where $\hat a = (\hat x + i \hat p)/\sqrt 2$.

\begin{figure}[h]
\centering
\includegraphics[scale=.28]{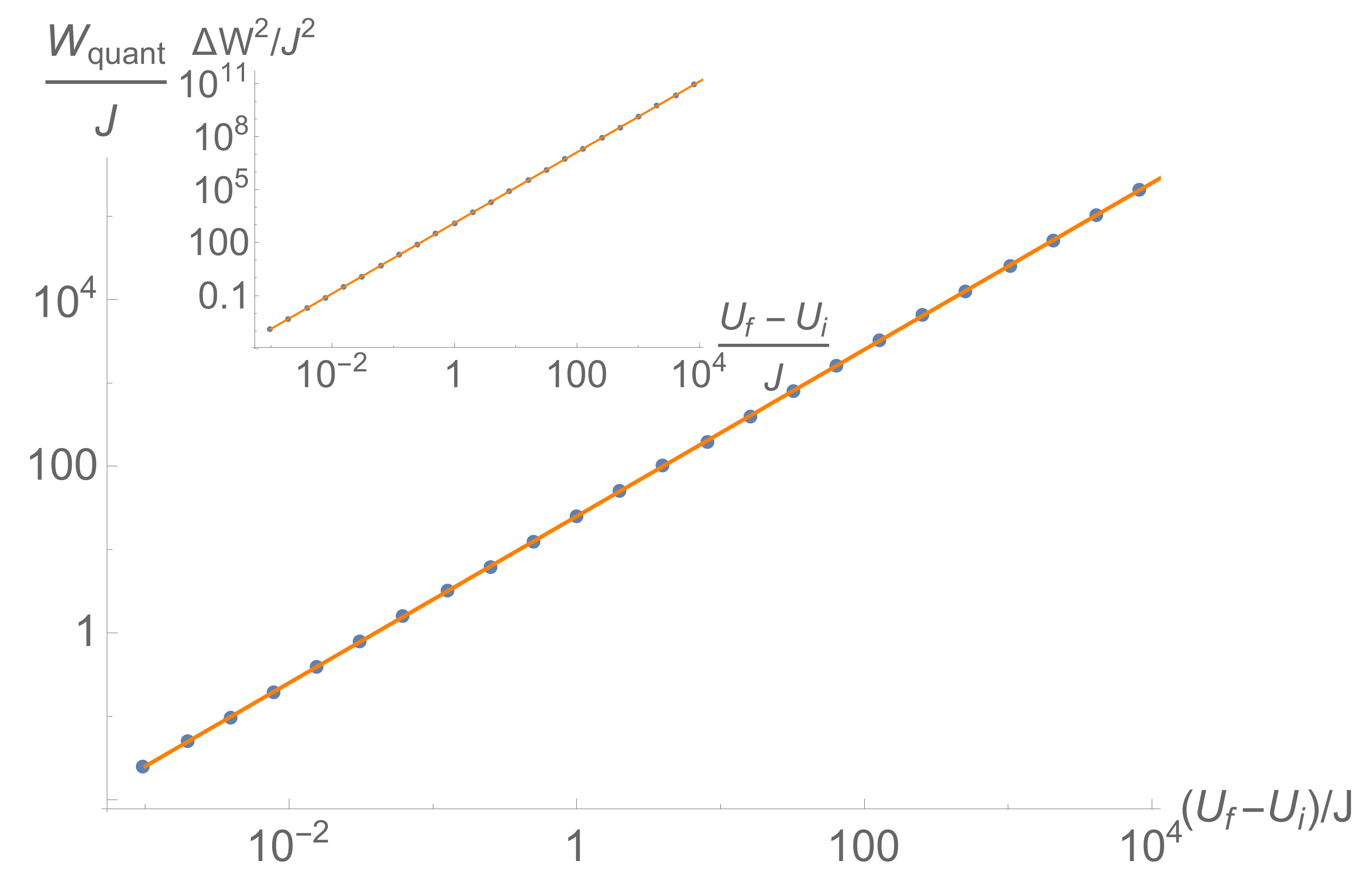}
\includegraphics[scale=.5]{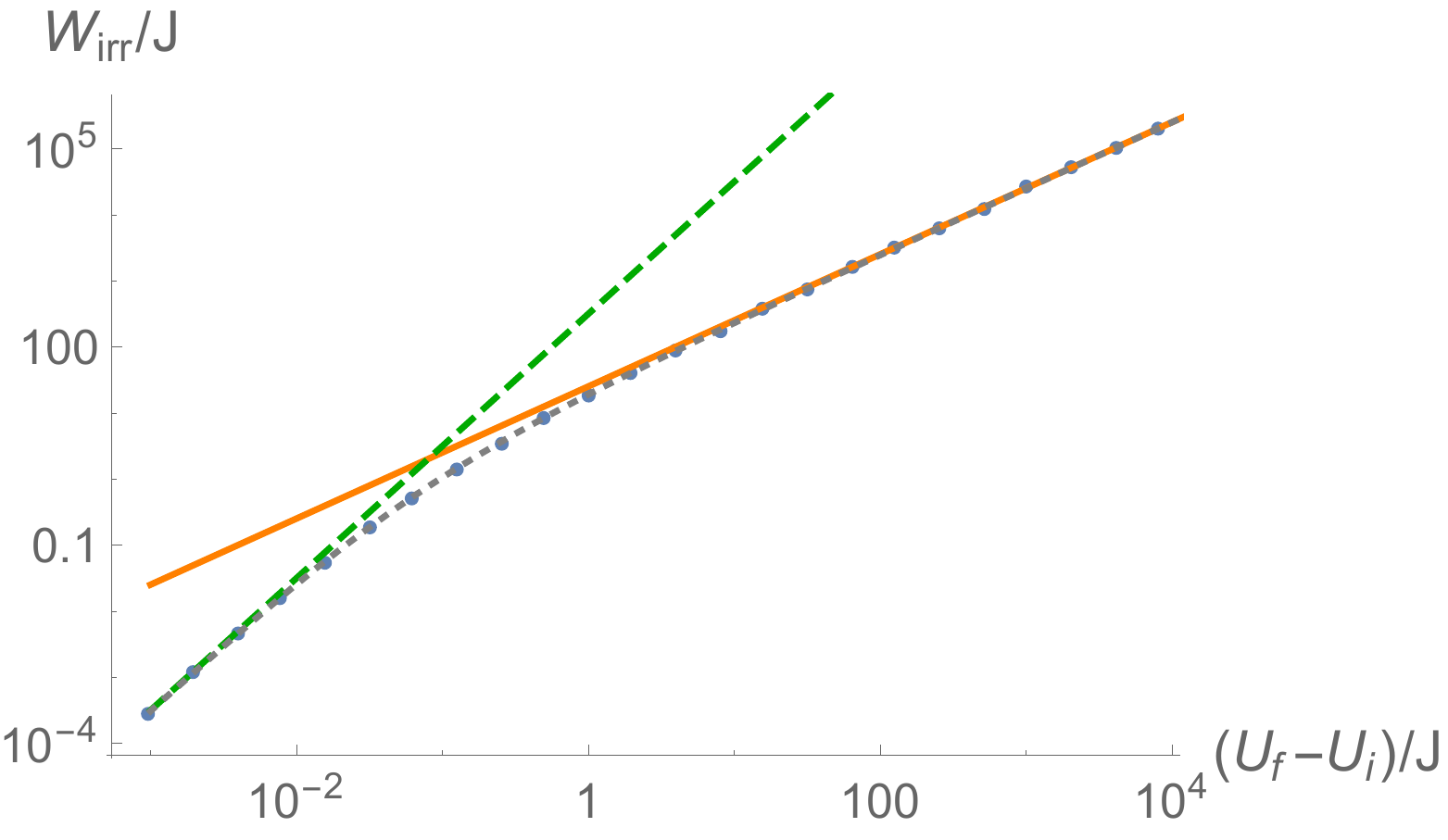} 
\caption{(Color online) Numerical (blue dots) and analytical (solid line) log-log plots of the quantum part of the average work (top, main), its variance (inset) and the irreversible work (bottom) against $(U_f-U_i)/J$, obtained by changing $U$ instantaneously from $U_i=0$, where the system is in the Rabi regime, to different values of $U_f$ in the range $10^{-3}J\div 10^4J$, including all the regimes for $N=100$. In the bottom panel, the dotted line fitting the numerical points corresponds to the QHO analytical prediction, the green (dashed) and the orange (solid) lines correspond to the limits $(U_f-U_i)/J$ much smaller and much larger than 1, respectively [see Eqs.\eqref{eq:wirr1} and \eqref{W_irr_Uigg1}].}
\label{fig:inst_quench_Rabi_all_Uf_LogLog}
\end{figure}

In the following we will need the ground state expectation values 
\begin{align}
\braket{\hat{x}^2}&=\frac{J}{\omega_p}
\label{x^2}\\
\braket{\hat{p}^2}&=\frac{\omega_p}{4J}
\end{align}
and using the Gaussian properties of the ground state, we obtain
\begin{equation}
\braket{\hat{x}^4}=3\braket{\hat{x}^2}.
\label{gaussian_moment}
\end{equation}
and similarly for other high order moments.

Within this framework, knowing the eigenstates of the QHO, we are able to compare both numerical and analytical results of the statistics of work which we now define. Suppose that we prepare a quantum system in the ground state $\ket{\psi_0}$ of an initial Hamiltonian  $\hat{H}_i$ with energy $E_0$. The Hamiltonian is then changed in time, not necessarily in an adiabatic fashion, reaching at time $\tau$ the Hamiltonian $\hat{H}_f$ with eigenvalues and eigenstates: $\{\tilde E_q,\ket{\tilde \psi_q}\}$. The change in the Hamiltonian induces an evolution operator that maps the initial state into $\ket{\psi(\tau)}$.
Then, the probability density function of the work done on the system is:
\begin{equation}
P(W)=\sum_q |\braket{\tilde \psi_q|\psi (\tau)}|^2\delta(W-\tilde E_q+E_0).
\end{equation}
A similar distribution can be analogously defined for an arbitrary initial state and for non unitary evolutions.

The average work done for the quench in a finite time, is then obtained as the first moment of $P(W)$:
\begin{equation}
\braket{W}=\braket{\psi(\tau)|\hat{H}_f|\psi(\tau)}-\braket{\psi_0|\hat{H}_i|\psi_0}.
\label{work_finite_time}
\end{equation}
The variance of the work, defined as
\begin{align}
\Delta W^2&=\braket{W^2}-\braket{W}^2,
\end{align}
with $\braket{W^2}=\braket{\psi(\tau)|(\hat{H}_f-E_0)^2|\psi(\tau)}$,  is useful because it gives information about the fluctuations of the work.

Thanks to the Jarzynski relation, it holds the relation $\braket{W}\geq \Delta F$, where $\Delta F=\tilde E_0 -E_0$ is the final-initial ground state energy difference. Since the equality holds in case of an adiabatic process, in the following we study the irreversible work $W_{irr}=\braket{W}-\Delta F$ which measures the amount of wasted work during the transformation.

\section{Instantaneous Quench}
We start our analysis with an instantaneous quench in which we vary either the self-interaction energy $U$ or the tunneling $J$. Under this assumption, $\ket{\psi(\tau)}=\ket{\psi_0}$ and the expectation value of \eqref{work_finite_time} is reduced to the evaluation on the ground state of the initial Hamiltonian, hence
\begin{equation}
\braket{W}=\braket{\psi_0|(\hat{H}_f-\hat{H}_i)|\psi_0}=\braket{\psi_0|\hat{H}_f|\psi_0}-E_0.
\end{equation}

 We evaluate the work done on the system going through all the regimes by keeping fixed $U_i$ and changing $U_f$ and vice versa. By using the mapping to the QHO as shown in Eq.~\eqref{QHO_Hamiltonian} and the results in Eqs.~(\ref{x^2}-\ref{gaussian_moment}), we find the following analytical results for the average work done on the system, its variance and the irreversible work
\begin{align}
\braket{W}&=\Delta U\frac{N}{2}\left[\left(\frac{N}{2}-1\right)+\frac{J}{\omega_i}\right]
\label{W_analytical_change_U}\\
\Delta W^2&=\frac{J^2}{\omega_i^2}\frac{N^2}{2}\Delta U^2
\label{varW_analytical_change_U}\\
W_{irr}&=\frac{N}{2}\frac{J}{\omega_i}\Delta U - \frac{\omega_f- \omega_i}{2}.
\label{Wirr_analytical_change_U}
\end{align}
where $\Delta U =U_f-U_i$ and $\omega_i,\omega_f$ are the initial and final plasma frequencies.

It is important to notice that the average work can be written as 
\begin{equation}
\braket{W}=W_{class}+W_{quant},
\end{equation}
where we want to stress the fact that the average work has a classical constant part $W_{class}=\Delta U\frac{N}{2}\left(\frac{N}{2}-1\right)$ and a quantum part affected by the ground state quantum fluctuations, related to the average square of the population imbalance
\begin{equation}
W_{quant}=\Delta U\frac{N}{2}\braket{\hat{x}^2}=\Delta U \braket{\hat{n}^2},
\label{quantum_work}
\end{equation}
where $\hat{n} =\hat n_L-\hat n_R$ is the population imbalance.

 \begin{figure}[t]
\centering
\includegraphics[scale=.4]{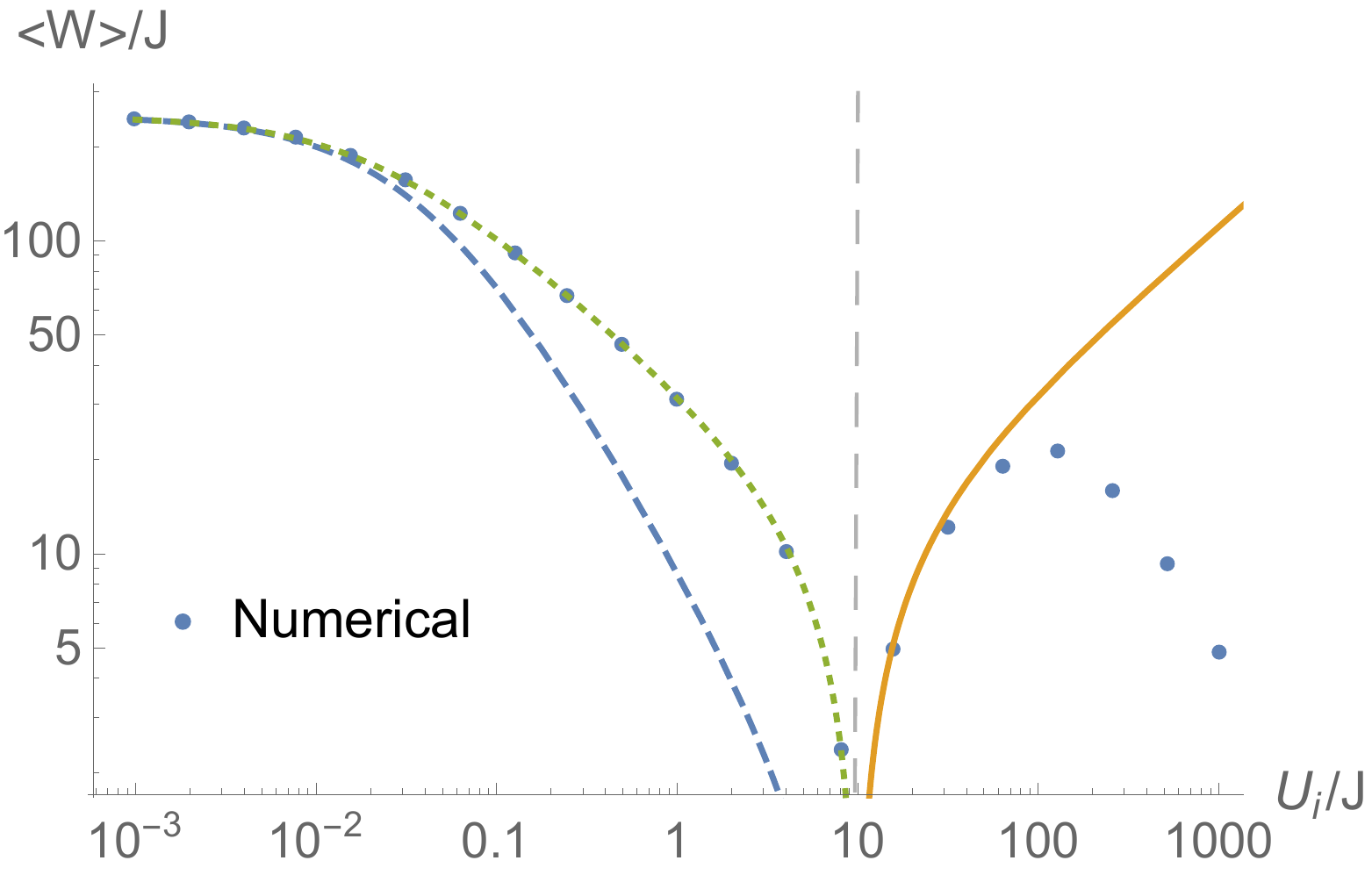}
\includegraphics[scale=.41]{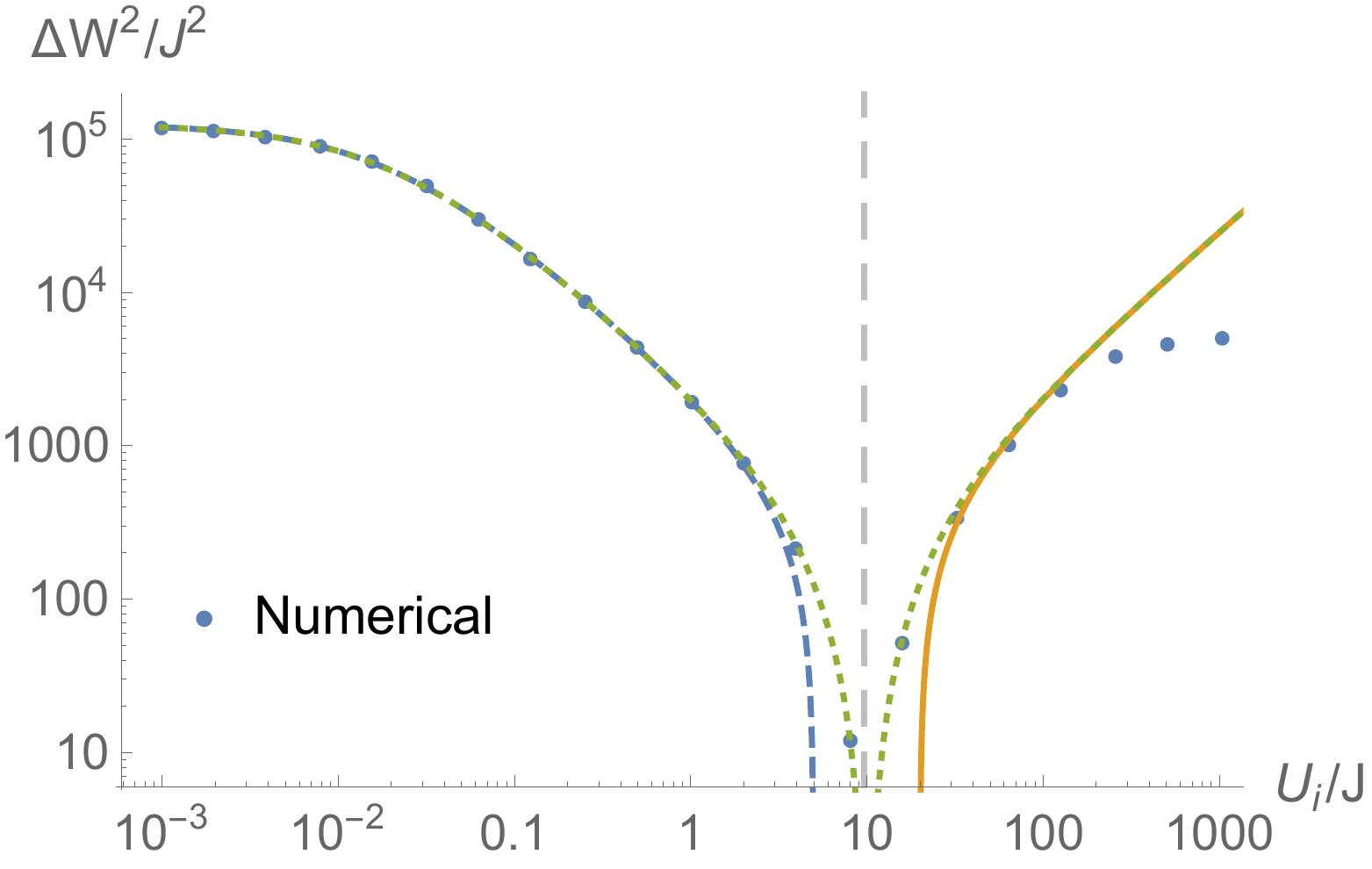}
\includegraphics[scale=.4]{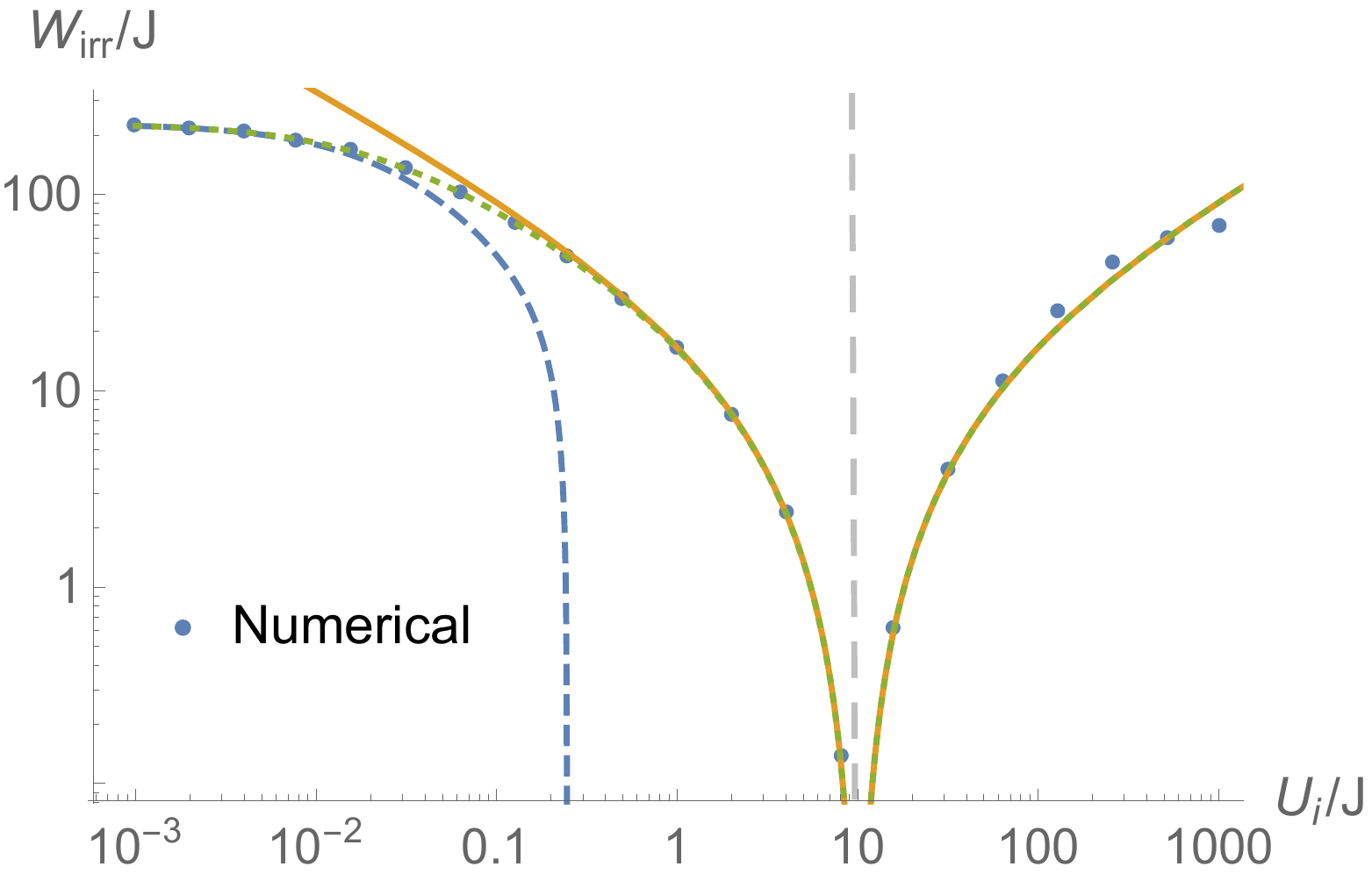}
\caption{(Color online) Analytical (green dotted lines) and numerical (blue dots) log-log plots of the quantum part of the average work (top), its variance (middle) and the irreversible work (bottom) vs $U_i$, obtained by changing $U$ instantaneously from different values of $U_i$ in the range $10^{-3}J\div 10^3J$, including all the regimes, to a value $U_f=10 J$, where the system is in the Josephson regime. We plot the absolute value of the work since it is negative for $U_i\geq U_f$. We set  $N=100$.
The blue (dashed) and orange (solid) lines represent respectively the analytical formulas obtained for the limiting cases of $\frac{U_i N}{2J}\ll 1$ and $\frac{U_i N}{2J}\gg 1$.}
\label{fig:inst_quench_all_Josephson_Ui_LogLog}
\end{figure}

In Fig. \ref{fig:inst_quench_Rabi_all_Uf_LogLog} we show the results of the quantum part of the average work, the variance and the irreversible work obtained by varying $U$, going from a fixed initial value $U_i$ where the system is in the Rabi regime, to different final values $U_f$ belonging to the three regimes.

As expected from the analytical expressions (\ref{W_analytical_change_U}-\ref{Wirr_analytical_change_U}), for the values used in Fig.~\ref{fig:inst_quench_Rabi_all_Uf_LogLog}, i.e. $U_i=0$ implying $\omega_i=2J$, we obtain $W_{quant}\approx \frac{N}{4} U_f$ and $\Delta W^2 \approx \frac{N^2}{8}U_f^2$. For the irreversible work the gray dotted line fitting the numerical points is given by the simplified form 
\begin{equation}
W_{irr}=\frac{NU_f}{4}+ J-J\sqrt{\frac{U_f N}{2J}+1}
\end{equation}
obtained from Eq.~\eqref{Wirr_analytical_change_U} with $U_i=0$.

 For this quantity we analyzed two limiting cases: $\frac{U_fN}{2J}\ll 1$ and $\frac{U_fN}{2J}\gg 1$. In the first scenario, by expanding the square root term up to the second order, i.e. $\sqrt{\frac{U_f N}{2J}+1}\simeq 1+\frac{U_f N}{4J}-\frac{U_f^2 N^2}{32 J^2}$, we get $W_{irr}\simeq \frac{N^2}{32J}U_f^2$, represented in Fig.~\ref{fig:inst_quench_Rabi_all_Uf_LogLog} by the green line. On the other hand, when $U_f$ increases and $\frac{U_fN}{2J}\gg 1$, the dominant term is the linear one, and we get $W_{irr}\simeq \frac{N}{4}U_f$ (orange line).

Analogously, in Fig.~\ref{fig:inst_quench_all_Josephson_Ui_LogLog} we study the case in which we keep fixed the final parameter $U_f$ and vary the initial one $U_i$ going through every of the three regimes of the bosonic Josephson junction.
Analogously to the previous case, for the average work, we consider only its quantum component \eqref{quantum_work}.

As expected, the analytical results (green dotted lines)  given by Eqs. \eqref{quantum_work}, \eqref{varW_analytical_change_U}, \eqref{Wirr_analytical_change_U} fit the numerical ones (blue dots) for $U_i< 10^3J$, because for larger values of $U_i$, with the parameters used here, the system is in the Fock regime. This limitation of the analytical approach is due to the fact that in this regime the Holstein-Primakoff approximation does not work anymore and the eigenstates of the initial Hamiltonian can not be described as the ones of the QHO.
In order to show the results on a log-log graphic, we considered the absolute values of the analysed quantities, since for $U_i>U_f$, i.e. at the right of the gray dotted line in the graphics, the average work has a negative value, corresponding to work extraction.
Furthermore, as done for the previous case in which we change $U_f$, we analyse the limiting cases for $\frac{U_i N}{2J}\gg 1$ and $\frac{U_i N}{2J}\ll 1$.

For $\frac{U_i N}{2J}\ll 1$ we obtain\\
\begin{align}
W_{quant}\simeq&\Delta U \frac{N J}{4J+U_i N}\\
\Delta W^2\simeq&(U_f^2 - 2 U_f U_i) \frac{N^2}{8} \frac{2}{N U_i/(2 J) + 1}\label{varW_series}\\
W_{irr}\simeq&\Delta U \frac{N}{4}\frac{1}{1+U_i N/4J}
\label{eq:wirr1}\\
&-\left[\omega_f -2J\left(1+\frac{U_i N}{4J}-\frac{U_i^2N^2}{32J^2}\right)\right]/2\nonumber,
\end{align}
where once again we use the expansion $$\sqrt{\frac{U_i N}{2J}+1}\simeq 1+\frac{U_iN}{4J}-\frac{U_i^2N^2}{32J^2}$$ in series up to the second order, and in \eqref{varW_series} we considered $U_i\ll U_f$.
These behaviours are shown by the blue lines in Fig.~\ref{fig:inst_quench_all_Josephson_Ui_LogLog}.

On the other hand, in the limit for $\frac{U_i N}{2J}\gg 1$, by using the approximation $\sqrt{\frac{U_i N}{2J}+1}\simeq\sqrt{\frac{U_i N}{2J}}$, we obtained the following analytical results for the examined quantities:
\begin{align}
W_{quant}\simeq&\frac{\Delta U}{4} \frac{2J N}{U_i}\\
\Delta W^2\simeq&-\frac{U_f N J}{4}-\Delta U\frac{NJ}{4} \\
W_{irr}\simeq&\frac{\Delta U}{4}\sqrt{\frac{2JN}{U_i}}-\sqrt{\frac{NJ}{2}}\left(\sqrt{U_f}-\sqrt{U_i}\right),
\label{W_irr_Uigg1}
\end{align}
where in \eqref{W_irr_Uigg1} we used the approximation $\frac{U_f N}{2J}\gg 1$, since the final state of the system is in the Josephson regime.

\begin{figure}[t]
\centering
\includegraphics[scale=.55]{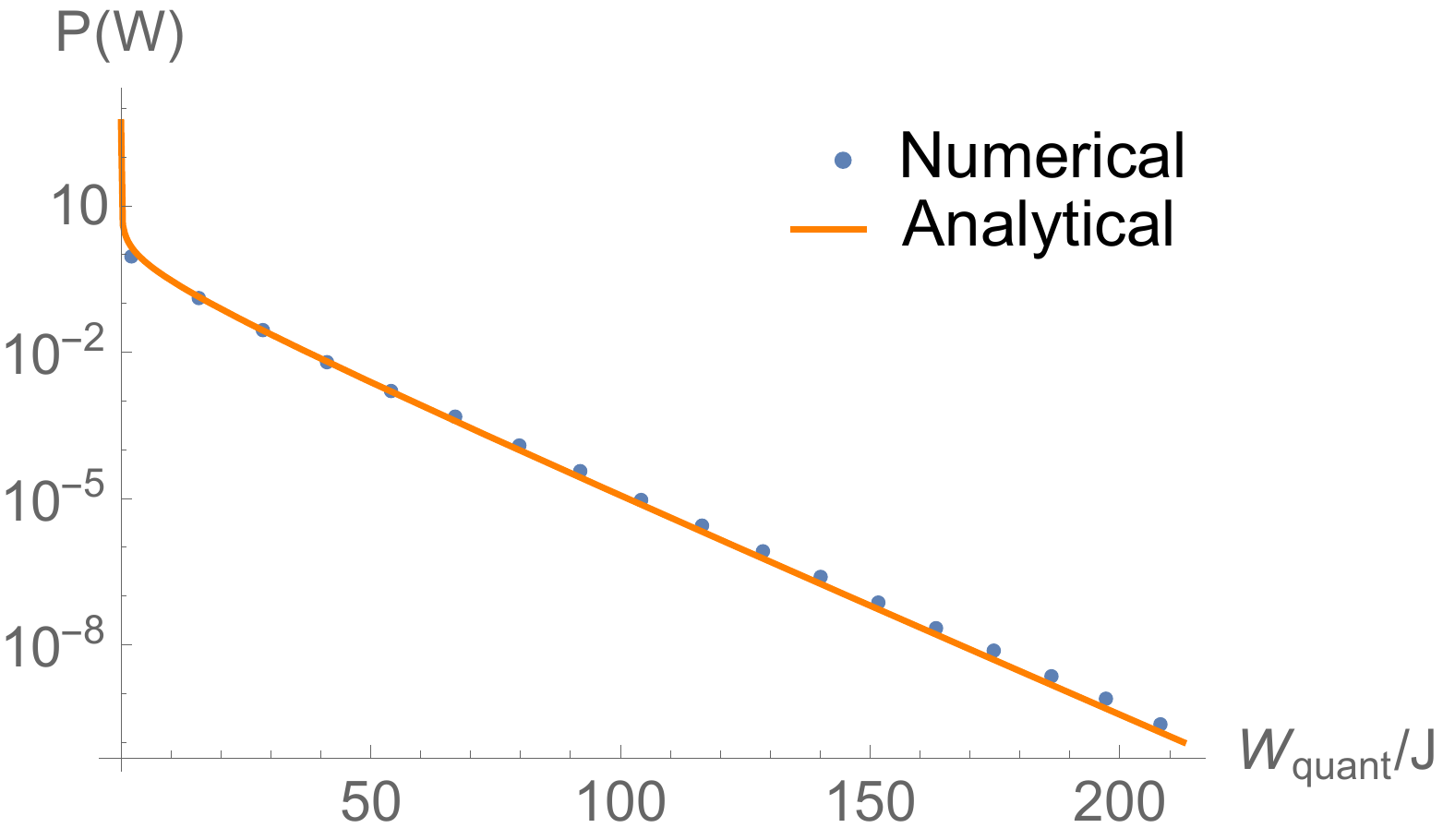} 
\caption{(Color online) Semi-log plot of the probability distribution of the work calculated analytically by using \eqref{quant_dist} and numerically, for a sudden quench from $U_i=0$ to $U_f=0.1 J$ and $N=100$.}
\label{fig:pd_quench}
\end{figure}

So far we have limited our analysis to the first two moments of work and the irreversible work. The full distribution of work can calculated in a similar way. As shown in Eq.~\eqref{quantum_work}, the quantum part of the work is proportional to the square of the population imbalance. Since this quantity is approximately Gaussian in the Rabi and Josephson regime, we expect $W_{quant}$ to be distributed according to an exponential function:
\begin{equation}
\label{quant_dist}
P(W_{quant}) =\sqrt{ \frac{N}{\pi \sigma W_{quant}}} \exp\left[-W/\sigma\right]
\end{equation}
where $\sigma=J \Delta U N/\omega_i$.
The relation \eqref{quant_dist} works quite well in the Josephson regime as shown in Fig.~\ref{fig:pd_quench}.
As it can be noticed from this plot, for higher values of the work, the analytical and numerical results present a progressive slight shift. This is probably due to the fact that in the Holstein-Primakoff approximation we are neglecting higher-order terms, hence the spacing between the energy levels in the Bose-Hubbard model may not be exactly the same as in the QHO.

\section{Finite time transformations}
We now turn to a transformation in which we vary the work parameter $U$  in a finite time $\tau$. As we saw in the previous sections, the properties of the bosonic Josephson junction are well captured by the QHO away from the Fock regime. We thus expect that even for the dynamics such mapping still holds. In our analysis, we compare the numerical results obtained for the work fluctuations by using the Bose-Hubbard model with the semi-analytical results obtained from the approach of Ford et al. \cite{Ford} for the evolution of the ground state in a QHO with a time-dependent frequency (see Appendix \ref{Appendix} for the detailed calculations).
Previous works have investigated the work distribution of a QHO, for the case of a linear ramp for the squared frequency \cite{DeffnerLutz}, i.e. $\omega^2(t)=\omega_0^2-(\omega_0^2-\omega_1^2)t/\tau$, and for a generic ramp \cite{Ford}.

\begin{figure}[t]
\centering
\includegraphics[scale=.4]{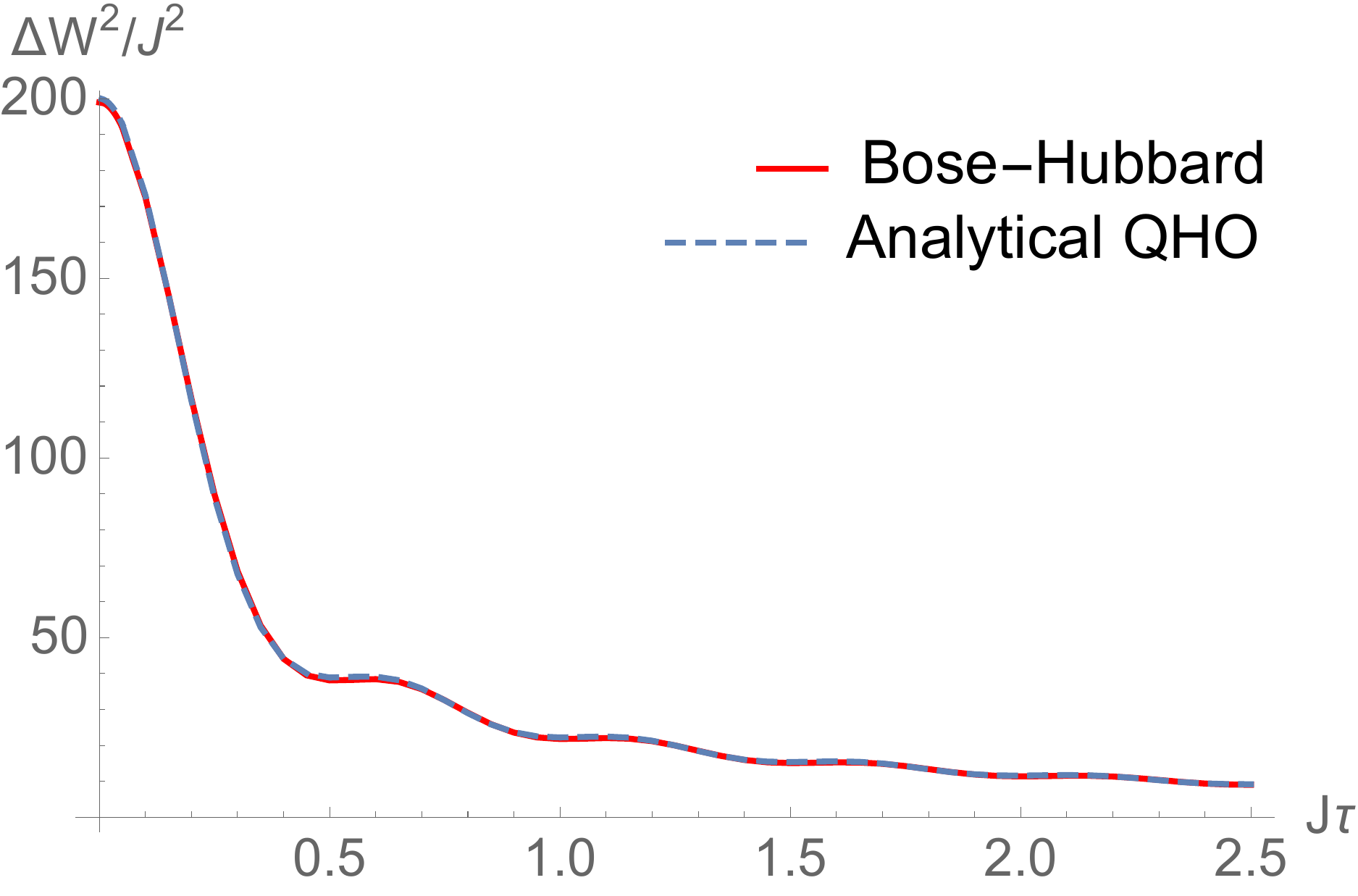}
\includegraphics[scale=.4]{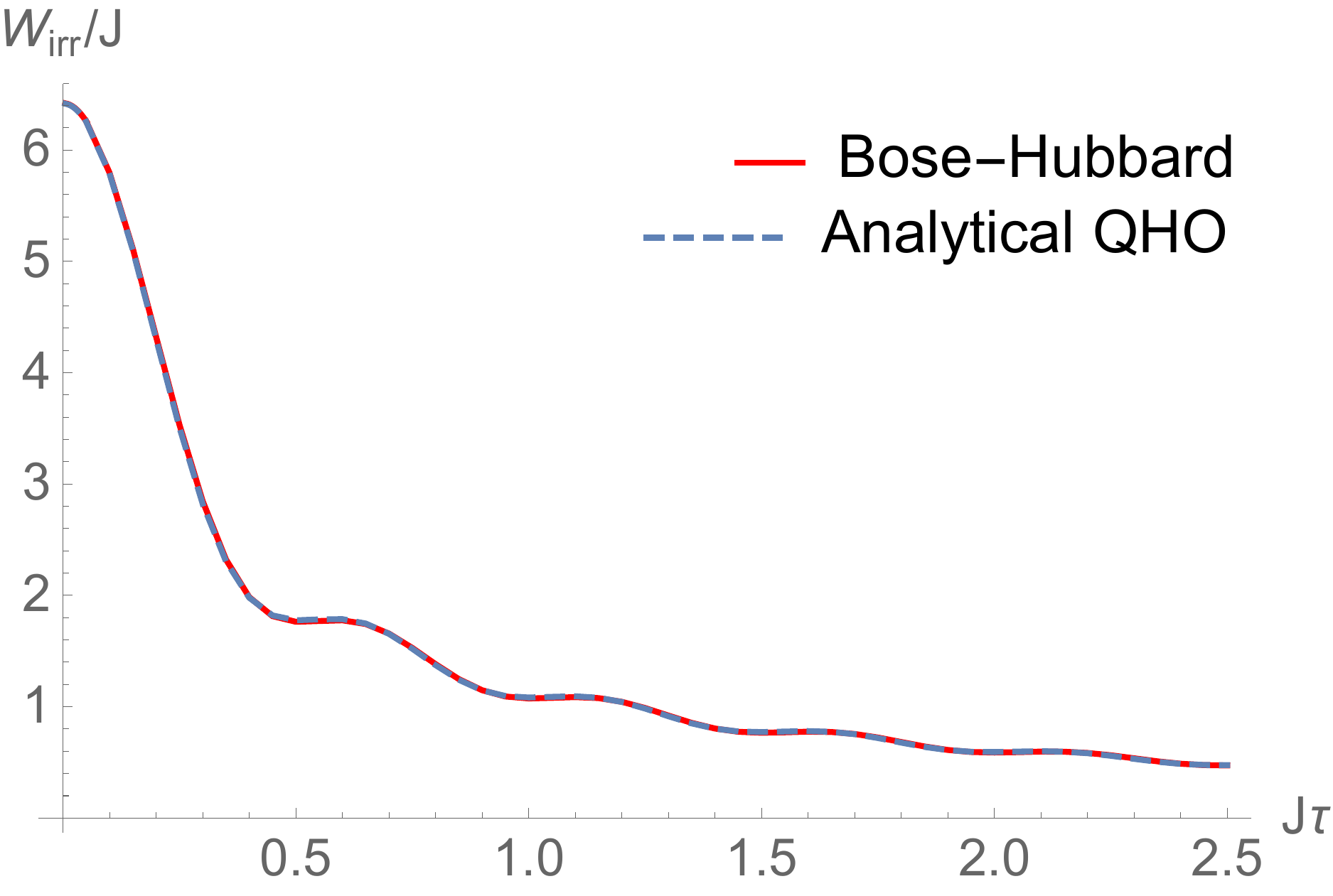}
\caption{(Color online) Analytical (dashed) and numerical (solid) plots of variance of the work (top) and the irreversible work (bottom) vs $\tau$, obtained by changing $U$ from $U_i=0$ to $U_f=0.2J$ with $N=200$.}
\label{fig:finite_time_Ui0_Uf0.2}
\end{figure}

We start our analysis with a linear ramp for $U(t)$:
\begin{equation}
\label{rampU}
U_{lin}(t)=U_i+(U_f-U_i)\frac{t}{\tau}.
\end{equation}
The results for the variance of work and irreversible work are shown in Fig.~\ref{fig:finite_time_Ui0_Uf0.2}. In order to compare the numerical results with the time evolution of the QHO, we notice that since the plasma frequency squared is a linear function of $U(t)$, for the ramp in Eq.~\eqref{rampU}, we are considering the same case of Ref. \cite{DeffnerLutz}.
The results show that both irreversible work and variance of work decay with the ramp duration $\tau$. This is analogous to the analysis in Ref. \cite{DeffnerLutz} with the adiabaticity parameter $Q$. Moreover we observe oscillations in both quantities as a function of $\tau$. These can be associated with parametric time oscillations of the variance of the population imbalance around the variance of the instantaneous ground state. Thus, the irreversible work is directly related to squeezing and anti-squeezing of the population imbalance in time. Such conjecture is confirmed in Appendix \ref{Appendix}.

Furthermore, having a semi-analytical form of the transition probability $p_{q,0}^{\tau}$ \eqref{transition_probability_ford1}, we obtain both numerical and analytical results for the probability distribution of the work, defined as
\begin{equation}
P(W)=\sum_q|p_{q,0}^{\tau}|^2\delta(W-q\omega_i).
\label{probability_distribution_ford}
\end{equation}

Even in this case, we obtain a shift between numerical and analytical results similar to the one obtained for a sudden quench, shown in Fig.~\ref{fig:pd_quench}.

\section{Optimal control}
It is natural to expect that for a given duration $\tau$, the irreversible work should depend on the actual time dependence of the self-interaction $U(t)$.
The aim of this section is to find the best ramp $U(t)$ that minimises $W_{irr}$ for fixed $\tau$. Previous attempts to reduce irreversible work in quantum harmonic oscillators \cite{Kosloff2009,GalveLutz} and systems within the linear response regime \cite{BonancaDeffner} have been reported. Our goal is a standard optimal control problem \cite{krotov1995global} which we approach considering two types of chopped basis: the first a linear ramp plus a  truncated Fourier expansion, similar to Ref. \cite{CRAB}, and a polynomial.

For these functions we impose the boundary conditions $U(0)=U_i$ and $U(\tau)=U_f$ and we enforce the plasma frequency $\omega(t)=2J\sqrt{\frac{U(t)N}{2J}+1}$ to be real for every $t$.
We optimized the free parameters of every kind of ramp and compared the results of the irreversible work with the case of the linear ramp $U_{lin}(t)$, Eq.~\eqref{rampU}.

In a first attempt, we use one ramp from each class with four parameters of which, given the boundary conditions, two are free.
The first one is a linear ramp with two sinusoidal terms, having the form
\begin{equation}
U_{LCS}(t)=A_0+A_1\cos\left(\frac{\pi t}{\tau}\right)+B_1\sin\left(\frac{\pi t}{\tau}\right)+C_1\frac{t}{\tau}
\end{equation}
with $A_0=U_i-A_1$ and $C_1=U_f-U_i+2A_1$
in which we optimize the free parameters $A_1$ and $B_1$. The frequency of the oscillating terms is chosen to have at least one oscillation during the ramp.
The second kind of ramp analysed is a cubic polynomial:
\begin{equation}
U_{c}(t)=A_0+A_1\frac{t}{\tau}+A_2\left(\frac{t}{\tau}\right)^2+A_3\left(\frac{t}{\tau}\right)^3,
\end{equation}
where the free parameters $A_2$ and $A_3$ are optimized. For this case the boundary conditions impose $A_0=U_i$ and $A_1=U_f-U_i-A_2-A_3$.

For comparison we consider also other ramps with four free parameters. In the first one we add two frequencies to the ramp $U_{LCS}(t)$, hence we consider $A_1$, $B_1$, $A_2$ and $B_2$ as free parameters in
\begin{align}
U_{2LCS}(t)=&A_0+A_1\cos\left(\frac{\pi t}{\tau}\right)+B_1\sin\left(\frac{\pi t}{\tau}\right)+\nonumber\\
&A_2\cos\left(\frac{2\pi t}{\tau}\right)+B_2\sin\left(\frac{2\pi t}{\tau}\right)+C_1\frac{t}{\tau}.
\end{align}

The last ramp we consider is a quintic polynomial:
\begin{align}
U_q(t)=&A_0+A_1\frac{t}{\tau}+A_2\left(\frac{t}{\tau}\right)^2+A_3\left(\frac{t}{\tau}\right)^3\nonumber\\
&+A_4\left(\frac{t}{\tau}\right)^4+A_5\left(\frac{t}{\tau}\right)^5.
\end{align}
In each of these cases, the condition of reality imposed on $\omega(t)$ gives a restriction on the possible values of one of the free parameters, depending on the values of the other ones.
The choice of the parameters range analysed is done on the basis of both efficiency and stability, by observing the dependence of the irreversible work on the free parameters which, for the ansatz having two free parameters, can be represented graphically as in Fig.~\ref{fig:LCS_Wirr_vs_A1&B1_tau0,1}.
\begin{figure}[t]
\includegraphics[scale=.28]{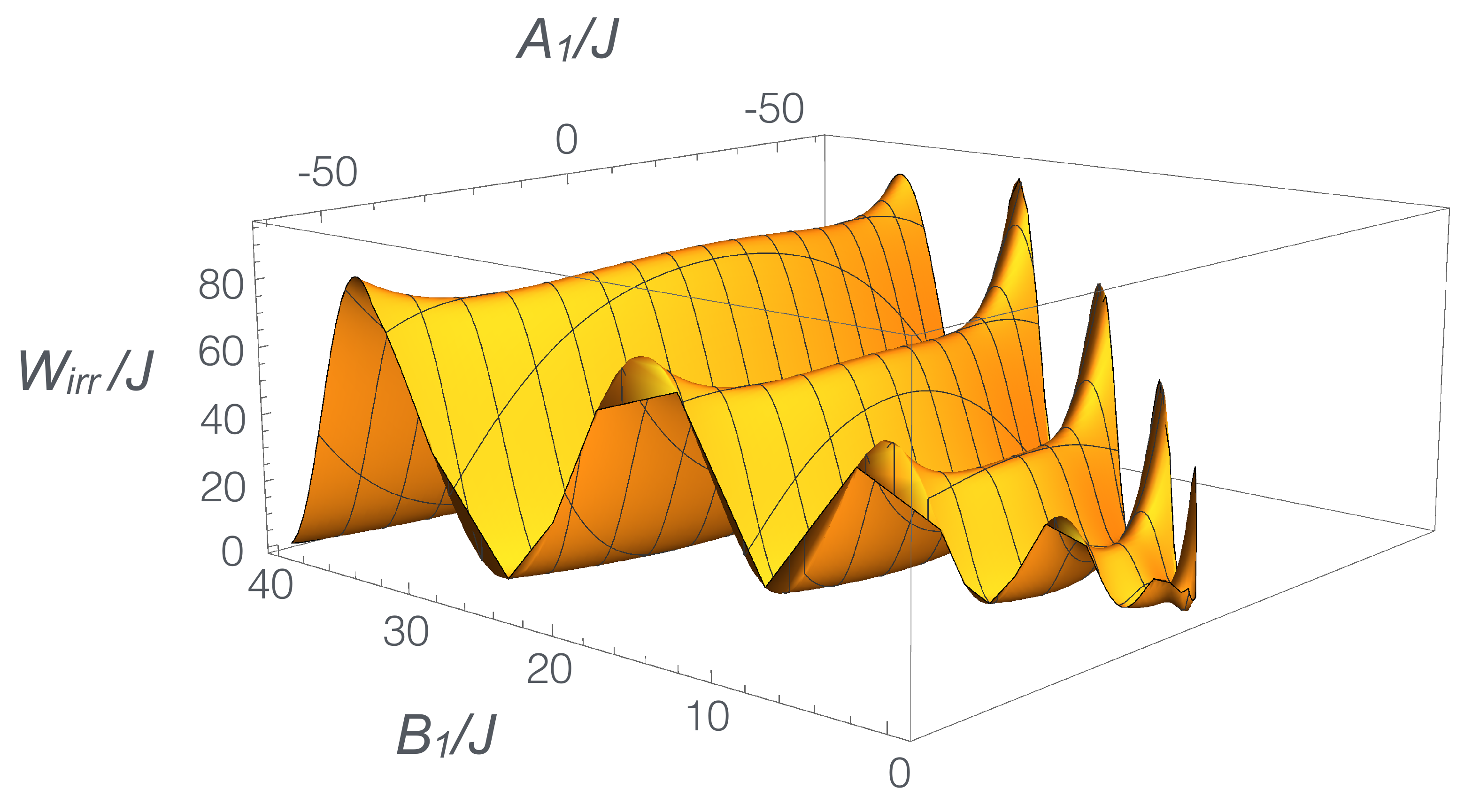} 
\caption{(Color online) Irreversible work vs the parameters $A_1$ and $B_2$ calculated for the ansatz $U_{LCS}(t)$ (Linear+Cos+Sin ramp) for the values $U_i=0.2J$, $U_f=0.8J$ and $\tau=0.1/J$.}
\label{fig:LCS_Wirr_vs_A1&B1_tau0,1}
\end{figure}
The plot of the irreversible work versus the free parameters for the ramp $U_{LSC}(t)$ in Fig.~\ref{fig:LCS_Wirr_vs_A1&B1_tau0,1} shows  oscillations of the irreversible work, whose amplitudes increase for larger values of the parameter $B_1$.
Nevertheless, this kind of considerations derived from a graphical representation is hard to extend to the case of more than two free parameters, but we expect a similar potential landscape \cite{Rabitz}.

\begin{figure}[b]
\includegraphics[scale=.5]{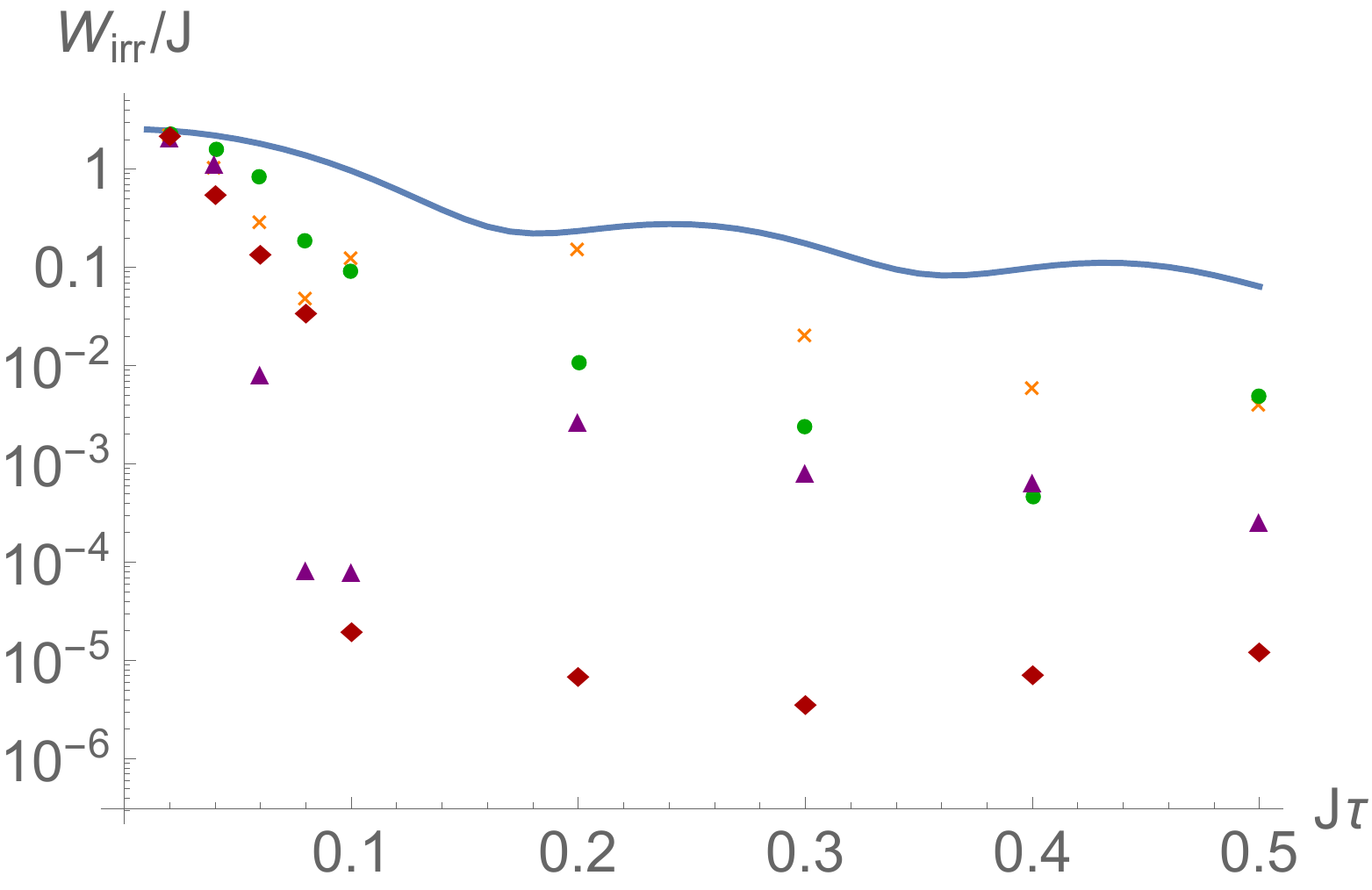}
\caption{(Color online) Semi-log plot of the irreversible work vs $\tau$ evaluated with the optimal parameters of every ansatz, for values $U_i=0.2J$ and $U_f=0.8J$.
The blue line corresponds to the linear ramp, the orange crosses to the linear+Sin+Cos ramp, the triangles to the ramp $U_{2LCS}(t)$, the green points and the diamonds respectively to the cubic and quintic ones.}
\label{fig:Wirr_vs_tau_all}
\end{figure}

The optimized results of the irreversible work obtained for every ramp as a function of the ramp duration $\tau$ are compared in Fig.~\ref{fig:Wirr_vs_tau_all} and reported in Fig.~\ref{table_opt_par} for convenience.
With each of these ramps we obtain a substantial decrease in the dissipated work with respect to the linear ramp for every $\tau$. The most efficient optimizations are obtained with the four-parameters ramps $U_{2LCS}(t)$ and  $U_q(t)$, for which the irreversible work reaches a value smaller than $10^{-4}J$ after a time $\tau=0.1/J$.
For the other ramps the value of the dissipated work is always larger and approaches zero for much higher values of $\tau$. In particular, for smaller values of the duration of the quench, i.e. $0<J\tau<0.08$, the ramp $U_{LCS}$ is more efficient than the cubic, but for $0.08<J\tau<0.3$,  the best optimization is granted by the cubic and the quintic, which reach a zero value of the dissipated work respectively from $\tau=0.08/J$ and $\tau=0.1/J$.

On the basis of these results regarding the efficiency of the optimization process, we analysed the stability of the optimal parameters obtained for every ramp. 
In order to do that, for every ansatz, we associate a relative percentage error to each parameter, we evaluate the work done on the system for random variations of the parameters inside the range given by the errors, and we consider the average and standard deviation of the irreversible work.

\begin{figure}[t]
\includegraphics[scale=.5]{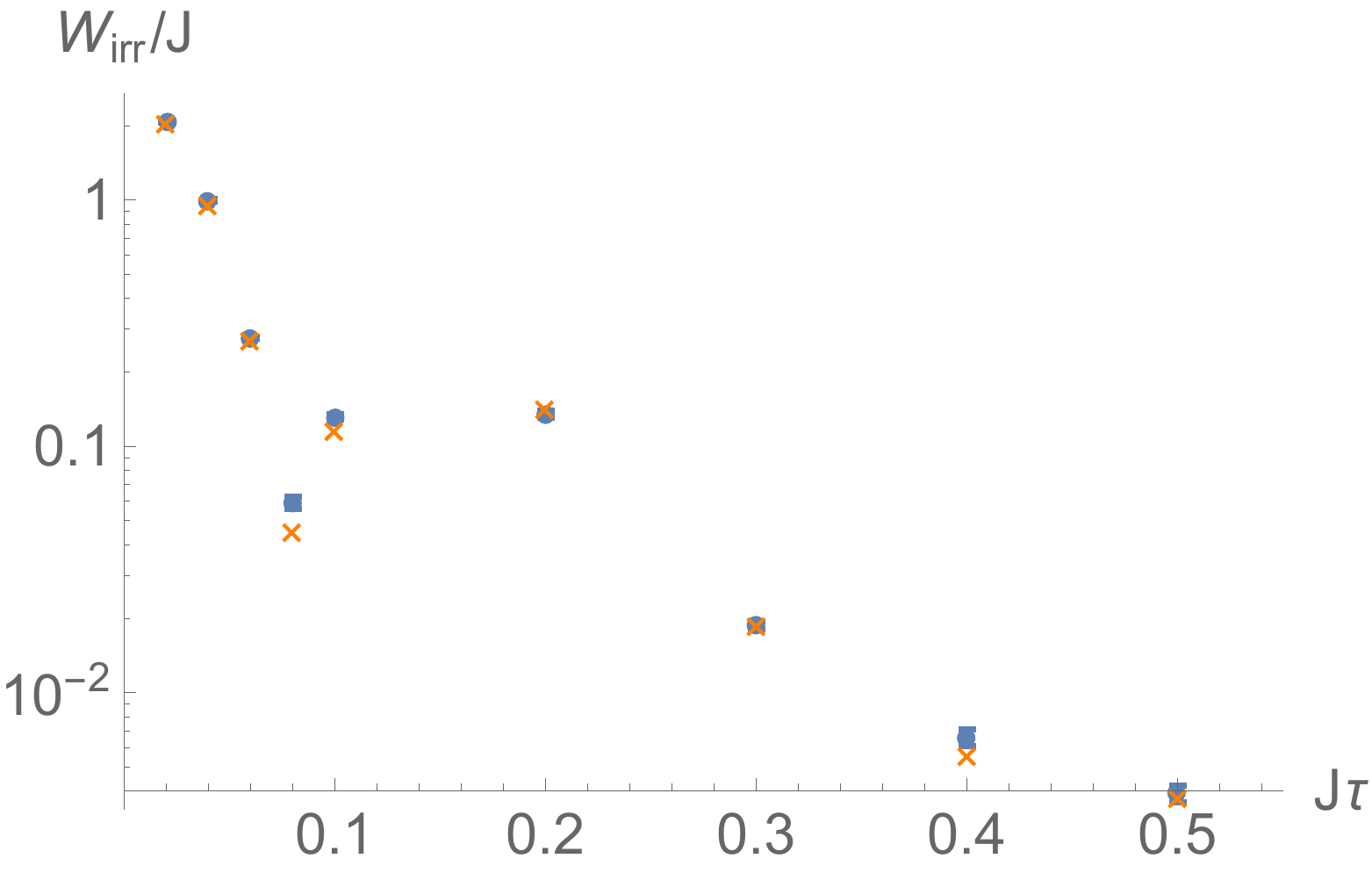}
\caption{(Color online) Plot of the irreversible work vs the duration of the quench $\tau$, evaluated for the exact optimized parameters (orange crosses) and as an average for random fluctuations of these parameters up to the $15\%$ and $5\%$ of the optimal values for $A_1$ and $B_1$ respectively (blue dots). These results are obtained for $U_i=0.2J$ and $U_f=0.8J$.}
\label{fig:LCS_Semilog_Wirr_err_A15_B5}
\end{figure}

For the case examined above, i.e. $U_i=0.2J$ and $U_f=0.8J$, we obtain the maximum stability for the ansatz $U_{LCS}$, where variations up to the $20\%$ of the optimal parameters give variations on the irreversible work between the $0.2\%$ and the $8\%$, except for the cases of $\tau=0.08/J$ and $\tau=0.1/J$ for which the variations are higher, respectively of the $13.6\%$ and the $11.4\%$. The results of this tolerance analysis are shown in Fig.~\ref{fig:LCS_Semilog_Wirr_err_A15_B5}.
For the cubic ramp, in order to obtain the same kind of results gained for the ramp $U_{LCS}$, we consider fluctuations of the parameters up to the $5\%$ of their value, obtaining variations in the irreversible work between the $0.5\%$ and the $14\%$, with higher peaks of the $61.5\%$ and the $40\%$ respectively for $\tau=0.1J$ and $\tau=0.3J$. The enhancement in the stability obtained with the ansatz $U_{LCS}$ is due to the fact that, as shown in Fig.~\ref{fig:LCS_Wirr_vs_A1&B1_tau0,1}, variations of $A_1$ tend to leave the irreversible work in a minimum, hence the major contribution to changes in the value of the work is given by $B_1$. On the other hand, for the cubic ramp, the minima of the oscillations cross different values of both $A_2$ and $A_3$, hence both the oscillations contribute to the variations of the irreversible work. 
Although the ansatz $U_{2LCS}$ gives the best efficiency, for the values of $U$ analysed, it is the most unstable. It would probably be possible to minimize the work analysing a different range of parameters, reducing the efficiency of the optimization in order to enhance its stability, but the lack of a graphical representation for the case of four free parameters analysis makes it harder to find a stable range.

\begin{table*}[t]
  \begin{tabular}{| c | rrl | rrrrl | rrl | rrrrl |}
    \hline
    \multirow{2}{*}{\backslashbox[0.7cm]{$\tau$}{\quad}} &
      \multicolumn{3}{c|}{$U_{LCS}$} &
      \multicolumn{5}{c|}{$U_{2LCS}$} &
      \multicolumn{3}{c|}{$U_c$} &
      \multicolumn{5}{c|}{$U_q$} \\
   & $A_1$ & $B_1$ & $W_{irr}$ & $A_1$ & $B_1$ & $A_2$ & $B_2$ & $W_{irr}$ & $A_2$ & $A_3$ & $W_{irr}$ & $A_2$ & $A_3$ & $A_4$ & $A_5$ & $W_{irr}$\\
    \hline
 0.02 &
 3.8 & 0.2 & 2.04 &
 3.6 & 0.4 & 0.2 & 0.0 & 2.01 &
 -8.0 & 5.4 & 2.29 &
  -15.8 & 0.6 & 16.2 & -7.4 & 2.13\\
  
 0.04 &
 3.2 & 0.0 & 9.47 $\times 10^{-1}$& 
 21.0 & 16.0 & -6.4 & -11.0 & 1.09 &
 -8.0 & 5.4 & 1.63 &
 25.0 & -14.0 & -21.4 & -26.0 & 5.36$\times 10^{-1}$\\
 
 0.06 &
 3.2  & 0.0 & 2.68 $\times 10^{-1}$&
 18.0 & 15.0 & 3.0 & -7.8 & 7.93 $\times 10^{-3}$& 
 -8.0 & 5.6 & 8.28 $\times 10^{-1}$ &
 19.6 & 19.0 & -20.2 & -26.0 & 1.31$\times 10^{-1}$\\
 
 0.08 &
 2.2 & -0.2 & 4.45 $\times 10^{-2}$ &
 1.6 & 10.4 & 4.2 & -1.4 & 7.96 $\times 10^{-5}$&
 -8.0 & 5.6 & 1.90 $\times 10^{-1}$&
 2.6 & 21.0 & -2.2 & -26.0 & 3.29 $\times 10^{-2}$\\
 
 0.10	&
 1.4 & -0.2 & 1.15 $\times 10^{-1}$&
-15.0 & 19.8 & 3.6 & -3.8 & 7.73 $\times 10^{-5}$&
 -7.6 & 5.4 & 1.84 $\times 10^{-3}$&
 4.8 & -14.2 & 8.2 & 1.2 & 1.90 $\times 10^{-5}$\\
 
 0.20 &
 0.2 & -0.2 & 1.41 $\times 10^{-1}$ &
 0.2 & 0.0 & -0.6 & 0.8 & 2.62 $\times 10^{-3}$&
 -0.4 & 0.6 & 1.09 $\times 10^{-2}$&
 6.8 & 21.2 & -16.8 & -17.0 & 6.67 $\times 10^{-6}$\\
 
 0.30 &
 0.0 & -0.2 & 1.86 $\times 10^{-2}$ &
 3.8 & 3.2 & 0.0 & -2.6 & 7.86 $\times 10^{-4}$ &
 0.0 & 0.6 & 2.36 $\times 10^{-3}$&
 17.8 & -14.0 & 13.8 & -20.4 & 3.44 $\times 10^{-6}$\\
 
 0.40 &
 0.2 & -0.2 & 5.49 $\times 10^{-3}$ &
 -2.8 & -0.4 & -3.6 & 2.4 & 6.23 $\times 10^{-4}$&
 -0.8 & 1.2 & 4.59 $\times 10^{-4}$&
 10.8 & -6.6 & 2.8 & -7.8 & 6.87 $\times 10^{-6}$\\
 
 0.50 &
 0.0 & -0.2 & 3.71 $\times 10^{-3}$ &
 -2.4 & 4.0 & 0.0 & -1.2 & 2.51 $\times 10^{-4}$&
 0.4 & 0.2 & 4.80 $\times 10^{-3}$&
 -1.0 & 14.0 & -20.0 & 7.0 & 1.19 $\times 10^{-5}$\\
    \hline
  \end{tabular}
  \caption{Values of the exact optimal parameters and the irreversible work obtained with these for each ramp, for different values of $\tau$, considering a quench from $U_i=0.2J$ to $U_f=0.8J$. All quantities are in units of $J$.}
  \label{table_opt_par}
\end{table*}

We also examine transitions in different regimes, going from an initial value $U_i=0.8J$ in the Josephson regime to a final one $U_f=40J$ in the Fock regime.
In this case, the most efficient ansatz are $U_{LCS}$ and $U_{2LCS}$, whose results of the irreversible work are the same from $\tau=0.02/J$ and for which a zero dissipated work is reached at $\tau=0.04/J$.
It is observed that with the cubic and the quintic the stability is enhanced, although the efficiency is lower, i.e. with the quintic ramp we find a dissipated work equal to zero at $\tau=0.1/J$ and for the cubic this is obtained at $\tau=0.4/J$, hence in this case a good compromise between optimization efficiency and stability would be the quintic ramp.

For the two cases examined we have found different results regarding the efficiency and the stability of the optimization, and this can be probably due to the fact that different boundary conditions give different constraints to the values that the parameters can assume. For this reason, we do not expect the kind of parameters landscape such as the one represented in Fig.~\ref{fig:LCS_Wirr_vs_A1&B1_tau0,1} to be the same in that range of parameter values for transitions with different values of $U_i$ and $U_f$.

\section{Conclusion}
In this paper we have analysed the fluctuations of the work done on an ensemble of ultracold atoms in a two-site Bose-Hubbard model. We have carefully shown analytical predictions for the first two moments of work and  the irreversible work for instantaneous quenches. In this regime we have predicted that the probability distribution of work is well described by an exponential function in agreement with that of a quantum harmonic oscillator.

For finite-time ramps, we have analysed the case of a linear ramp in time demonstrating oscillations of the irreversible work that are synchronous with the squeezing oscillations of the population imbalance distribution. Finally we have used simple optimal control techniques to minimise the irreversible work to negligible values. This result might have applications in the realisation of quantum thermal machines with ultracold atoms and in the quest to maximise their efficiencies. 

It is natural to expect further decrease of the irreversible work using a larger chopped basis for the time-dependent ramp. Moreover, one could use more sophisticated schemes as the one in Ref. \cite{CRAB}, in which the frequencies of the oscillating terms are chosen random, or the Krotov's method \cite{krotov1995global}. Our analysis is therefore a starting point for a more systematic study.

Finally, let us discuss possible experimental verifications of our analysis. The two-site Bose-Hubbard model can be realised in different setups with ultracold atoms in double well potentials, in atomic condensates with two species, and in self-organised condensates in optical cavities \cite{GatiPRL,schumm2005matter,Schmiedmayer2014,brennecke2013real}. It is worth to stress the fact that \eqref{quantum_work} is an interesting result because it shows that we can reconstruct the statistics of the work, at least the first two moments, experimentally by measuring only one observable, i.e. the square of the population imbalance, rather than doing two measurements on the energy of the system or by coupling the system to an external quantum probe. Similar experiments could be carried out in nuclear magnetic resonance quadrupolar systems in which spin squeezing has been recently observed \cite{Auccaise,Auccaise2}.

\acknowledgements
The authors thank F. Cataliotti, M. Fattori, I. Mekhov, and J. Sherson for invaluable discussions. This work is supported by the John Templeton Foundation (grant ID 43467), the EU Collaborative Project TherMiQ (Grant Agreement 618074).

\appendix
\section{Work distribution of a QHO}
\label{Appendix}

In this section we revise the dynamics of the parametrically driven quantum harmonic oscillator and how to calculate the work distribution.
Let us assume the QHO to be driven with Hamiltonian Eq.~\eqref{QHO_Hamiltonian} from an initial frequency $\omega(0)=\omega_i$ to a final one $\omega(\tau)=\omega_1$.
An initial energy eigenfunction 
of a QHO evolves into \cite{Husimi01041953,Ford}
\begin{eqnarray}
&&\psi_n(t)=\frac{1}{2^n n!}\left(\frac{\omega_i}{\pi g_-(t)}\right)^{1/4}H_n\left(\sqrt{\frac{\omega_i}{g_-(t)}}x\right)\cdot\\
&&\exp\biggl[-\frac{i g_0(t)+\omega_i}{2\pi g_-(t)}x^2
-i\left(n+\frac{1}{2}\right)\int_0^t \frac{\omega_i}{m g_-(t')}dt'\biggr]\nonumber,
\label{evolved_QHO_state_Ford}
\end{eqnarray}
where the $H_n$ are the Hermite polynomials and the functions $g_+(t)$, $g_-(t)$ and $g_0(t)$ satisfy the differential equations
\begin{align}
\dot{g}_-(t)&=-2g_0(t)/m\\
\dot{g}_0(t)&=m\omega^2(t)g_-(t)-g_+(t)/m\\
\dot{g}_+(t)&=2m\omega^2(t)g_0(t)
\end{align}
with initial conditions $g_-(0)=1/m$, $g_0(0)=0$ and $g_+(0)=m\omega_0^2$.

Eq.~\eqref{evolved_QHO_state_Ford} can be used to calculate the position variance of the QHO initially in its ground state $n=0$:
\begin{align}
\braket{x^2}&=\int_{-\infty}^{\infty}|\psi_0(t)|^2x^2dx\nonumber\\
&=\left(\frac{\omega_i}{\pi g_-(t)}\right)^{\frac{1}{2}}\int_{-\infty}^{\infty}dx x^2 \exp\left[2\text{Re}\left(\frac{ig_0(t)-\omega_i}{2 g_-(t)}\right) x^2 \right]\nonumber\\
&=\frac{1}{2}\sqrt{\frac{\omega_i}{g_-(t)}}\left[2 \text{Re}\left(\frac{ig_0(t)-\omega_i}{2 g_-(t)}\right)\right]^{3/2} \label{variance_qho_ford}
\end{align}
where in the last passage we used the Gaussian integral $\int e^{-\lambda x^2} x^2 dx = \frac{1}{2}\sqrt{\pi}\lambda^{-3/2}$.
It can be noticed that in the case of an adiabatic ramp, since the evolved state at the time $\tau$ is an eigenstate of the final Hamiltonian, $\braket{x^2}$ will be the variance of the ground state for the final Hamiltonian.

\begin{figure}[t]
\centering
\includegraphics[scale=.35]{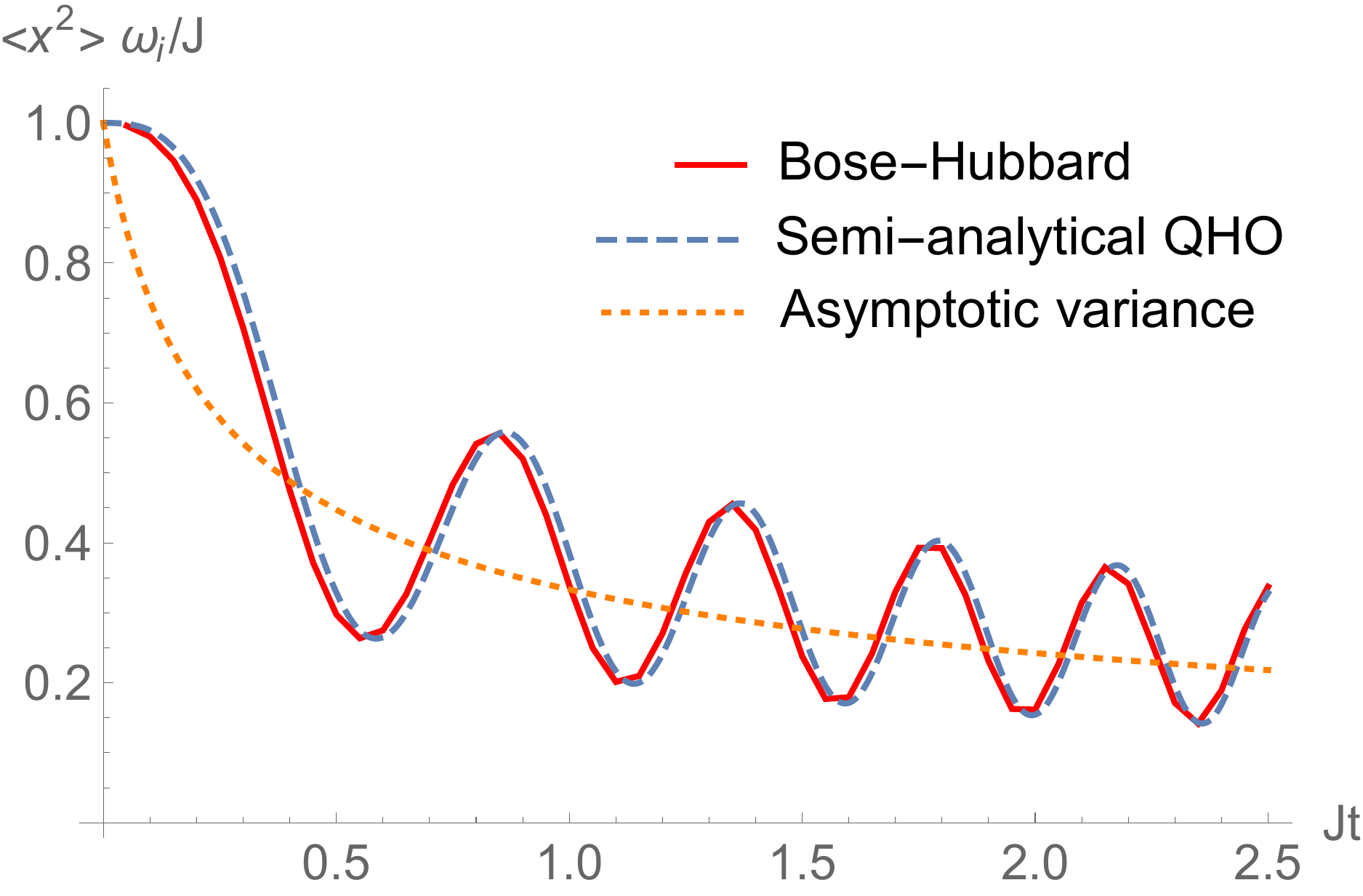} 
\caption{(Color online) Comparison of the plots of the variance in units of $J/\omega_i$ (where $\omega_i$ is the plasma frequency at $t=0$) vs time obtained for the cases of the numerical Bose-Hubbard model, the semi-analytical approach to the QHO and the asymptotic limit. The parameters used are $N=200$, $U_i=0$, $U_f=0.2 J$.}
\label{fig:variance_linear_ramp}
\end{figure}

In Fig.~\ref{fig:variance_linear_ramp} ,we compare the analytical result from \eqref{variance_qho_ford} with the one we obtained numerically by using a Trotter expansion of the evolution operator and to the instantaneous variance, defined as $\braket{x^2(t)}={J}/{\omega(t)}$. As the frequency $\omega(t)$ is increased, the system wavefunction tries to catch-up with the instantaneous value of the variance and start oscillating around it. The slower is the driving the smaller is the amplitude of these squeezing oscillations.

To calculate the probability distribution of the work we need the transition probability
\begin{align}
p_{q,0}=|\braket{\tilde \psi_q|\psi (\tau)}|^2 =\left|\int_{-\infty}^{\infty}dx\tilde{\psi}^*_q\psi_0(\tau)\right|^2.
\end{align}
where we set $n=0$ as we assume the initial state to be the ground state.
To this end,  the wavefunctions of the final Hamiltonian at $t=\tau$, are given by
\begin{equation}
\tilde{\psi}_q=\frac{1}{2^q q!}\left(\frac{m\omega_f}{\pi}\right)^{1/4}\exp\biggl[\frac{-m\omega_f x^2}{2}\biggr]H_q\left(\sqrt{m\omega_f}x\right).
\end{equation}

We thus obtain:
\begin{align}
p_{q,0}&=\biggl|\int_{-\infty}^{\infty}dx\tilde{\psi}_q\psi_0(\tau)\biggr|^2=\\
&=\biggl|\frac{1}{2^q q!}\left(\frac{m\omega_f}{\pi}\right)^{1/4}\left(\frac{\omega_i}{\pi g_-(t)}\right)^{1/4}\\
&\int_{-\infty}^{\infty}dx H_q\left(\sqrt{m\omega_f}x\right) \exp\biggl[\left(\frac{i g_0(t)-\omega_i}{2\pi g_-(t)}-\frac{m\omega_f}{2}\right)x^2\biggr]\biggr|^2\nonumber
\label{transition_probability_ford1}
\end{align}
Because of the parity, the only possible transitions that give a non-zero value for the integral above, are the ones for which the index $q$ is even, and in this case from \eqref{transition_probability_ford1} we obtain the result
\begin{align}
p_{q,0}=&\biggl|\frac{1}{2^q}\left(\frac{m\omega_f\omega_i}{g_-(t)}\right)^{1/4}\frac{1}{(q/2)!}\\
&\frac{[(ig_0(t)-\omega_i)/(2\pi g_-(t))-m\omega_f/2+\sqrt{m\omega_f}]^{q/2}}{[-(ig_0(t)-\omega_i)/(2\pi g_-(t))+m\omega_f/2]^{(q+1)/2}}\biggl|^2\nonumber,
\label{transition_probability_ford}
\end{align}
where we used the result
\begin{equation*}
\int \exp[\alpha x^2]H_q(\beta x) dx=\dfrac{q!\sqrt{\pi}}{(q/2)!}\dfrac{(\alpha+\beta)^{q/2}}{(-\alpha)^{(q+1)/2}}.
\end{equation*}

\bibliographystyle{apsrev4-1}
\bibliography{biblio}

\begin{thebibliography}{43}%
\makeatletter
\providecommand \@ifxundefined [1]{%
 \@ifx{#1\undefined}
}%
\providecommand \@ifnum [1]{%
 \ifnum #1\expandafter \@firstoftwo
 \else \expandafter \@secondoftwo
 \fi
}%
\providecommand \@ifx [1]{%
 \ifx #1\expandafter \@firstoftwo
 \else \expandafter \@secondoftwo
 \fi
}%
\providecommand \natexlab [1]{#1}%
\providecommand \enquote  [1]{``#1''}%
\providecommand \bibnamefont  [1]{#1}%
\providecommand \bibfnamefont [1]{#1}%
\providecommand \citenamefont [1]{#1}%
\providecommand \href@noop [0]{\@secondoftwo}%
\providecommand \href [0]{\begingroup \@sanitize@url \@href}%
\providecommand \@href[1]{\@@startlink{#1}\@@href}%
\providecommand \@@href[1]{\endgroup#1\@@endlink}%
\providecommand \@sanitize@url [0]{\catcode `\\12\catcode `\$12\catcode
  `\&12\catcode `\#12\catcode `\^12\catcode `\_12\catcode `\%12\relax}%
\providecommand \@@startlink[1]{}%
\providecommand \@@endlink[0]{}%
\providecommand \url  [0]{\begingroup\@sanitize@url \@url }%
\providecommand \@url [1]{\endgroup\@href {#1}{\urlprefix }}%
\providecommand \urlprefix  [0]{URL }%
\providecommand \Eprint [0]{\href }%
\providecommand \doibase [0]{http://dx.doi.org/}%
\providecommand \selectlanguage [0]{\@gobble}%
\providecommand \bibinfo  [0]{\@secondoftwo}%
\providecommand \bibfield  [0]{\@secondoftwo}%
\providecommand \translation [1]{[#1]}%
\providecommand \BibitemOpen [0]{}%
\providecommand \bibitemStop [0]{}%
\providecommand \bibitemNoStop [0]{.\EOS\space}%
\providecommand \EOS [0]{\spacefactor3000\relax}%
\providecommand \BibitemShut  [1]{\csname bibitem#1\endcsname}%
\let\auto@bib@innerbib\@empty
\bibitem [{\citenamefont {Jarzynski}(1997)}]{jarzynski1997nonequilibrium}%
  \BibitemOpen
  \bibfield  {author} {\bibinfo {author} {\bibfnamefont {C.}~\bibnamefont
  {Jarzynski}},\ }\href {\doibase 10.1103/PhysRevLett.78.2690} {\bibfield
  {journal} {\bibinfo  {journal} {Phys. Rev. Lett.}\ }\textbf {\bibinfo
  {volume} {78}},\ \bibinfo {pages} {2690} (\bibinfo {year}
  {1997})}\BibitemShut {NoStop}%
\bibitem [{\citenamefont {Crooks}(1999)}]{Crooks}%
  \BibitemOpen
  \bibfield  {author} {\bibinfo {author} {\bibfnamefont {G.~E.}\ \bibnamefont
  {Crooks}},\ }\href {\doibase 10.1103/PhysRevE.60.2721} {\bibfield  {journal}
  {\bibinfo  {journal} {Phys. Rev. E}\ }\textbf {\bibinfo {volume} {60}},\
  \bibinfo {pages} {2721} (\bibinfo {year} {1999})}\BibitemShut {NoStop}%
\bibitem [{\citenamefont {Talkner}\ \emph {et~al.}(2007)\citenamefont
  {Talkner}, \citenamefont {Lutz},\ and\ \citenamefont
  {H\"anggi}}]{TalknerPRE2007}%
  \BibitemOpen
  \bibfield  {author} {\bibinfo {author} {\bibfnamefont {P.}~\bibnamefont
  {Talkner}}, \bibinfo {author} {\bibfnamefont {E.}~\bibnamefont {Lutz}}, \
  and\ \bibinfo {author} {\bibfnamefont {P.}~\bibnamefont {H\"anggi}},\ }\href
  {\doibase 10.1103/PhysRevE.75.050102} {\bibfield  {journal} {\bibinfo
  {journal} {Phys. Rev. E}\ }\textbf {\bibinfo {volume} {75}},\ \bibinfo
  {pages} {050102} (\bibinfo {year} {2007})}\BibitemShut {NoStop}%
\bibitem [{\citenamefont {Campisi}\ \emph {et~al.}(2011)\citenamefont
  {Campisi}, \citenamefont {H\"anggi},\ and\ \citenamefont
  {Talkner}}]{CampisiRMP}%
  \BibitemOpen
  \bibfield  {author} {\bibinfo {author} {\bibfnamefont {M.}~\bibnamefont
  {Campisi}}, \bibinfo {author} {\bibfnamefont {P.}~\bibnamefont {H\"anggi}}, \
  and\ \bibinfo {author} {\bibfnamefont {P.}~\bibnamefont {Talkner}},\ }\href
  {\doibase 10.1103/RevModPhys.83.771} {\bibfield  {journal} {\bibinfo
  {journal} {Rev. Mod. Phys.}\ }\textbf {\bibinfo {volume} {83}},\ \bibinfo
  {pages} {771} (\bibinfo {year} {2011})}\BibitemShut {NoStop}%
\bibitem [{\citenamefont {Watanabe}\ \emph {et~al.}(2014)\citenamefont
  {Watanabe}, \citenamefont {Venkatesh}, \citenamefont {Talkner}, \citenamefont
  {Campisi},\ and\ \citenamefont {H\"anggi}}]{Watanabe}%
  \BibitemOpen
  \bibfield  {author} {\bibinfo {author} {\bibfnamefont {G.}~\bibnamefont
  {Watanabe}}, \bibinfo {author} {\bibfnamefont {B.~P.}\ \bibnamefont
  {Venkatesh}}, \bibinfo {author} {\bibfnamefont {P.}~\bibnamefont {Talkner}},
  \bibinfo {author} {\bibfnamefont {M.}~\bibnamefont {Campisi}}, \ and\
  \bibinfo {author} {\bibfnamefont {P.}~\bibnamefont {H\"anggi}},\ }\href
  {\doibase 10.1103/PhysRevE.89.032114} {\bibfield  {journal} {\bibinfo
  {journal} {Phys. Rev. E}\ }\textbf {\bibinfo {volume} {89}},\ \bibinfo
  {pages} {032114} (\bibinfo {year} {2014})}\BibitemShut {NoStop}%
\bibitem [{\citenamefont {Roncaglia}\ \emph {et~al.}(2014)\citenamefont
  {Roncaglia}, \citenamefont {Cerisola},\ and\ \citenamefont
  {Paz}}]{RoncagliaPRL}%
  \BibitemOpen
  \bibfield  {author} {\bibinfo {author} {\bibfnamefont {A.~J.}\ \bibnamefont
  {Roncaglia}}, \bibinfo {author} {\bibfnamefont {F.}~\bibnamefont {Cerisola}},
  \ and\ \bibinfo {author} {\bibfnamefont {J.~P.}\ \bibnamefont {Paz}},\ }\href
  {\doibase 10.1103/PhysRevLett.113.250601} {\bibfield  {journal} {\bibinfo
  {journal} {Phys. Rev. Lett.}\ }\textbf {\bibinfo {volume} {113}},\ \bibinfo
  {pages} {250601} (\bibinfo {year} {2014})}\BibitemShut {NoStop}%
\bibitem [{\citenamefont {{De Chiara}}\ \emph {et~al.}(2015)\citenamefont {{De
  Chiara}}, \citenamefont {Roncaglia},\ and\ \citenamefont
  {Paz}}]{DeChiaraPaz}%
  \BibitemOpen
  \bibfield  {author} {\bibinfo {author} {\bibfnamefont {G.}~\bibnamefont {{De
  Chiara}}}, \bibinfo {author} {\bibfnamefont {A.~J.}\ \bibnamefont
  {Roncaglia}}, \ and\ \bibinfo {author} {\bibfnamefont {J.~P.}\ \bibnamefont
  {Paz}},\ }\href {http://stacks.iop.org/1367-2630/17/i=3/a=035004} {\bibfield
  {journal} {\bibinfo  {journal} {New Journal of Physics}\ }\textbf {\bibinfo
  {volume} {17}},\ \bibinfo {pages} {035004} (\bibinfo {year}
  {2015})}\BibitemShut {NoStop}%
\bibitem [{\citenamefont {Talkner}\ and\ \citenamefont
  {H\"anggi}(2016)}]{TalknerHanggi2015}%
  \BibitemOpen
  \bibfield  {author} {\bibinfo {author} {\bibfnamefont {P.}~\bibnamefont
  {Talkner}}\ and\ \bibinfo {author} {\bibfnamefont {P.}~\bibnamefont
  {H\"anggi}},\ }\href {\doibase 10.1103/PhysRevE.93.022131} {\bibfield
  {journal} {\bibinfo  {journal} {Phys. Rev. E}\ }\textbf {\bibinfo {volume}
  {93}},\ \bibinfo {pages} {022131} (\bibinfo {year} {2016})}\BibitemShut
  {NoStop}%
\bibitem [{\citenamefont {Fusco}\ \emph {et~al.}(2014)\citenamefont {Fusco},
  \citenamefont {Pigeon}, \citenamefont {Apollaro}, \citenamefont {Xuereb},
  \citenamefont {Mazzola}, \citenamefont {Campisi}, \citenamefont {Ferraro},
  \citenamefont {Paternostro},\ and\ \citenamefont {De~Chiara}}]{Fusco}%
  \BibitemOpen
  \bibfield  {author} {\bibinfo {author} {\bibfnamefont {L.}~\bibnamefont
  {Fusco}}, \bibinfo {author} {\bibfnamefont {S.}~\bibnamefont {Pigeon}},
  \bibinfo {author} {\bibfnamefont {T.~J.~G.}\ \bibnamefont {Apollaro}},
  \bibinfo {author} {\bibfnamefont {A.}~\bibnamefont {Xuereb}}, \bibinfo
  {author} {\bibfnamefont {L.}~\bibnamefont {Mazzola}}, \bibinfo {author}
  {\bibfnamefont {M.}~\bibnamefont {Campisi}}, \bibinfo {author} {\bibfnamefont
  {A.}~\bibnamefont {Ferraro}}, \bibinfo {author} {\bibfnamefont
  {M.}~\bibnamefont {Paternostro}}, \ and\ \bibinfo {author} {\bibfnamefont
  {G.}~\bibnamefont {De~Chiara}},\ }\href {\doibase 10.1103/PhysRevX.4.031029}
  {\bibfield  {journal} {\bibinfo  {journal} {Phys. Rev. X}\ }\textbf {\bibinfo
  {volume} {4}},\ \bibinfo {pages} {031029} (\bibinfo {year}
  {2014})}\BibitemShut {NoStop}%
\bibitem [{\citenamefont {Bloch}\ \emph {et~al.}(2008)\citenamefont {Bloch},
  \citenamefont {Dalibard},\ and\ \citenamefont {Zwerger}}]{BlochRMP}%
  \BibitemOpen
  \bibfield  {author} {\bibinfo {author} {\bibfnamefont {I.}~\bibnamefont
  {Bloch}}, \bibinfo {author} {\bibfnamefont {J.}~\bibnamefont {Dalibard}}, \
  and\ \bibinfo {author} {\bibfnamefont {W.}~\bibnamefont {Zwerger}},\ }\href
  {\doibase 10.1103/RevModPhys.80.885} {\bibfield  {journal} {\bibinfo
  {journal} {Rev. Mod. Phys.}\ }\textbf {\bibinfo {volume} {80}},\ \bibinfo
  {pages} {885} (\bibinfo {year} {2008})}\BibitemShut {NoStop}%
\bibitem [{\citenamefont {Georgescu}\ \emph {et~al.}(2014)\citenamefont
  {Georgescu}, \citenamefont {Ashhab},\ and\ \citenamefont {Nori}}]{NoriRMP}%
  \BibitemOpen
  \bibfield  {author} {\bibinfo {author} {\bibfnamefont {I.~M.}\ \bibnamefont
  {Georgescu}}, \bibinfo {author} {\bibfnamefont {S.}~\bibnamefont {Ashhab}}, \
  and\ \bibinfo {author} {\bibfnamefont {F.}~\bibnamefont {Nori}},\ }\href
  {\doibase 10.1103/RevModPhys.86.153} {\bibfield  {journal} {\bibinfo
  {journal} {Rev. Mod. Phys.}\ }\textbf {\bibinfo {volume} {86}},\ \bibinfo
  {pages} {153} (\bibinfo {year} {2014})}\BibitemShut {NoStop}%
\bibitem [{\citenamefont {Lewenstein}\ \emph {et~al.}(2012)\citenamefont
  {Lewenstein}, \citenamefont {Sanpera},\ and\ \citenamefont
  {Ahufinger}}]{lewenstein2012ultracold}%
  \BibitemOpen
  \bibfield  {author} {\bibinfo {author} {\bibfnamefont {M.}~\bibnamefont
  {Lewenstein}}, \bibinfo {author} {\bibfnamefont {A.}~\bibnamefont {Sanpera}},
  \ and\ \bibinfo {author} {\bibfnamefont {V.}~\bibnamefont {Ahufinger}},\
  }\href {https://books.google.co.uk/books?id=Wpl91RDxV5IC} {\emph {\bibinfo
  {title} {Ultracold Atoms in Optical Lattices}}}\ (\bibinfo  {publisher} {OUP
  Oxford},\ \bibinfo {year} {2012})\BibitemShut {NoStop}%
\bibitem [{\citenamefont {Eisert}\ \emph {et~al.}(2015)\citenamefont {Eisert},
  \citenamefont {Friesdorf},\ and\ \citenamefont
  {Gogolin}}]{eisert2015quantum}%
  \BibitemOpen
  \bibfield  {author} {\bibinfo {author} {\bibfnamefont {J.}~\bibnamefont
  {Eisert}}, \bibinfo {author} {\bibfnamefont {M.}~\bibnamefont {Friesdorf}}, \
  and\ \bibinfo {author} {\bibfnamefont {C.}~\bibnamefont {Gogolin}},\
  }\href@noop {} {\bibfield  {journal} {\bibinfo  {journal} {Nature Physics}\
  }\textbf {\bibinfo {volume} {11}},\ \bibinfo {pages} {124} (\bibinfo {year}
  {2015})}\BibitemShut {NoStop}%
\bibitem [{\citenamefont {Polkovnikov}\ \emph {et~al.}(2011)\citenamefont
  {Polkovnikov}, \citenamefont {Sengupta}, \citenamefont {Silva},\ and\
  \citenamefont {Vengalattore}}]{PolkoRMP}%
  \BibitemOpen
  \bibfield  {author} {\bibinfo {author} {\bibfnamefont {A.}~\bibnamefont
  {Polkovnikov}}, \bibinfo {author} {\bibfnamefont {K.}~\bibnamefont
  {Sengupta}}, \bibinfo {author} {\bibfnamefont {A.}~\bibnamefont {Silva}}, \
  and\ \bibinfo {author} {\bibfnamefont {M.}~\bibnamefont {Vengalattore}},\
  }\href {\doibase 10.1103/RevModPhys.83.863} {\bibfield  {journal} {\bibinfo
  {journal} {Rev. Mod. Phys.}\ }\textbf {\bibinfo {volume} {83}},\ \bibinfo
  {pages} {863} (\bibinfo {year} {2011})}\BibitemShut {NoStop}%
\bibitem [{\citenamefont {Josephson}(1962)}]{josephson1962possible}%
  \BibitemOpen
  \bibfield  {author} {\bibinfo {author} {\bibfnamefont {B.~D.}\ \bibnamefont
  {Josephson}},\ }\href@noop {} {\bibfield  {journal} {\bibinfo  {journal}
  {Physics letters}\ }\textbf {\bibinfo {volume} {1}},\ \bibinfo {pages} {251}
  (\bibinfo {year} {1962})}\BibitemShut {NoStop}%
\bibitem [{\citenamefont {Javanainen}(1986)}]{javanainen1986oscillatory}%
  \BibitemOpen
  \bibfield  {author} {\bibinfo {author} {\bibfnamefont {J.}~\bibnamefont
  {Javanainen}},\ }\href {\doibase 10.1103/PhysRevLett.57.3164} {\bibfield
  {journal} {\bibinfo  {journal} {Phys. Rev. Lett.}\ }\textbf {\bibinfo
  {volume} {57}},\ \bibinfo {pages} {3164} (\bibinfo {year}
  {1986})}\BibitemShut {NoStop}%
\bibitem [{\citenamefont {Smerzi}\ \emph {et~al.}(1997)\citenamefont {Smerzi},
  \citenamefont {Fantoni}, \citenamefont {Giovanazzi},\ and\ \citenamefont
  {Shenoy}}]{Giovanazzi}%
  \BibitemOpen
  \bibfield  {author} {\bibinfo {author} {\bibfnamefont {A.}~\bibnamefont
  {Smerzi}}, \bibinfo {author} {\bibfnamefont {S.}~\bibnamefont {Fantoni}},
  \bibinfo {author} {\bibfnamefont {S.}~\bibnamefont {Giovanazzi}}, \ and\
  \bibinfo {author} {\bibfnamefont {S.~R.}\ \bibnamefont {Shenoy}},\ }\href
  {\doibase 10.1103/PhysRevLett.79.4950} {\bibfield  {journal} {\bibinfo
  {journal} {Phys. Rev. Lett.}\ }\textbf {\bibinfo {volume} {79}},\ \bibinfo
  {pages} {4950} (\bibinfo {year} {1997})}\BibitemShut {NoStop}%
\bibitem [{\citenamefont {Milburn}\ \emph {et~al.}(1997)\citenamefont
  {Milburn}, \citenamefont {Corney}, \citenamefont {Wright},\ and\
  \citenamefont {Walls}}]{Milburn}%
  \BibitemOpen
  \bibfield  {author} {\bibinfo {author} {\bibfnamefont {G.~J.}\ \bibnamefont
  {Milburn}}, \bibinfo {author} {\bibfnamefont {J.}~\bibnamefont {Corney}},
  \bibinfo {author} {\bibfnamefont {E.~M.}\ \bibnamefont {Wright}}, \ and\
  \bibinfo {author} {\bibfnamefont {D.~F.}\ \bibnamefont {Walls}},\ }\href
  {\doibase 10.1103/PhysRevA.55.4318} {\bibfield  {journal} {\bibinfo
  {journal} {Phys. Rev. A}\ }\textbf {\bibinfo {volume} {55}},\ \bibinfo
  {pages} {4318} (\bibinfo {year} {1997})}\BibitemShut {NoStop}%
\bibitem [{\citenamefont {Leggett}(2001)}]{LeggettRMP}%
  \BibitemOpen
  \bibfield  {author} {\bibinfo {author} {\bibfnamefont {A.~J.}\ \bibnamefont
  {Leggett}},\ }\href {\doibase 10.1103/RevModPhys.73.307} {\bibfield
  {journal} {\bibinfo  {journal} {Rev. Mod. Phys.}\ }\textbf {\bibinfo {volume}
  {73}},\ \bibinfo {pages} {307} (\bibinfo {year} {2001})}\BibitemShut
  {NoStop}%
\bibitem [{\citenamefont {Juli\'a-D\'{\i}az}\ \emph {et~al.}(2010)\citenamefont
  {Juli\'a-D\'{\i}az}, \citenamefont {Dagnino}, \citenamefont {Lewenstein},
  \citenamefont {Martorell},\ and\ \citenamefont {Polls}}]{JuliaDiaz}%
  \BibitemOpen
  \bibfield  {author} {\bibinfo {author} {\bibfnamefont {B.}~\bibnamefont
  {Juli\'a-D\'{\i}az}}, \bibinfo {author} {\bibfnamefont {D.}~\bibnamefont
  {Dagnino}}, \bibinfo {author} {\bibfnamefont {M.}~\bibnamefont {Lewenstein}},
  \bibinfo {author} {\bibfnamefont {J.}~\bibnamefont {Martorell}}, \ and\
  \bibinfo {author} {\bibfnamefont {A.}~\bibnamefont {Polls}},\ }\href
  {\doibase 10.1103/PhysRevA.81.023615} {\bibfield  {journal} {\bibinfo
  {journal} {Phys. Rev. A}\ }\textbf {\bibinfo {volume} {81}},\ \bibinfo
  {pages} {023615} (\bibinfo {year} {2010})}\BibitemShut {NoStop}%
\bibitem [{\citenamefont {Barzanjeh}\ and\ \citenamefont
  {Vitali}(2016)}]{Barzanjeh}%
  \BibitemOpen
  \bibfield  {author} {\bibinfo {author} {\bibfnamefont {S.}~\bibnamefont
  {Barzanjeh}}\ and\ \bibinfo {author} {\bibfnamefont {D.}~\bibnamefont
  {Vitali}},\ }\href {\doibase 10.1103/PhysRevA.93.033846} {\bibfield
  {journal} {\bibinfo  {journal} {Phys. Rev. A}\ }\textbf {\bibinfo {volume}
  {93}},\ \bibinfo {pages} {033846} (\bibinfo {year} {2016})}\BibitemShut
  {NoStop}%
\bibitem [{\citenamefont {Albiez}\ \emph {et~al.}(2005)\citenamefont {Albiez},
  \citenamefont {Gati}, \citenamefont {F\"olling}, \citenamefont {Hunsmann},
  \citenamefont {Cristiani},\ and\ \citenamefont {Oberthaler}}]{GatiPRL}%
  \BibitemOpen
  \bibfield  {author} {\bibinfo {author} {\bibfnamefont {M.}~\bibnamefont
  {Albiez}}, \bibinfo {author} {\bibfnamefont {R.}~\bibnamefont {Gati}},
  \bibinfo {author} {\bibfnamefont {J.}~\bibnamefont {F\"olling}}, \bibinfo
  {author} {\bibfnamefont {S.}~\bibnamefont {Hunsmann}}, \bibinfo {author}
  {\bibfnamefont {M.}~\bibnamefont {Cristiani}}, \ and\ \bibinfo {author}
  {\bibfnamefont {M.~K.}\ \bibnamefont {Oberthaler}},\ }\href {\doibase
  10.1103/PhysRevLett.95.010402} {\bibfield  {journal} {\bibinfo  {journal}
  {Phys. Rev. Lett.}\ }\textbf {\bibinfo {volume} {95}},\ \bibinfo {pages}
  {010402} (\bibinfo {year} {2005})}\BibitemShut {NoStop}%
\bibitem [{\citenamefont {Gati}\ and\ \citenamefont
  {Oberthaler}(2007)}]{GatiJPB}%
  \BibitemOpen
  \bibfield  {author} {\bibinfo {author} {\bibfnamefont {R.}~\bibnamefont
  {Gati}}\ and\ \bibinfo {author} {\bibfnamefont {M.~K.}\ \bibnamefont
  {Oberthaler}},\ }\href {http://stacks.iop.org/0953-4075/40/i=10/a=R01}
  {\bibfield  {journal} {\bibinfo  {journal} {Journal of Physics B: Atomic,
  Molecular and Optical Physics}\ }\textbf {\bibinfo {volume} {40}},\ \bibinfo
  {pages} {R61} (\bibinfo {year} {2007})}\BibitemShut {NoStop}%
\bibitem [{\citenamefont {Shin}\ \emph {et~al.}(2004)\citenamefont {Shin},
  \citenamefont {Saba}, \citenamefont {Pasquini}, \citenamefont {Ketterle},
  \citenamefont {Pritchard},\ and\ \citenamefont {Leanhardt}}]{Shin}%
  \BibitemOpen
  \bibfield  {author} {\bibinfo {author} {\bibfnamefont {Y.}~\bibnamefont
  {Shin}}, \bibinfo {author} {\bibfnamefont {M.}~\bibnamefont {Saba}}, \bibinfo
  {author} {\bibfnamefont {T.~A.}\ \bibnamefont {Pasquini}}, \bibinfo {author}
  {\bibfnamefont {W.}~\bibnamefont {Ketterle}}, \bibinfo {author}
  {\bibfnamefont {D.~E.}\ \bibnamefont {Pritchard}}, \ and\ \bibinfo {author}
  {\bibfnamefont {A.~E.}\ \bibnamefont {Leanhardt}},\ }\href {\doibase
  10.1103/PhysRevLett.92.050405} {\bibfield  {journal} {\bibinfo  {journal}
  {Phys. Rev. Lett.}\ }\textbf {\bibinfo {volume} {92}},\ \bibinfo {pages}
  {050405} (\bibinfo {year} {2004})}\BibitemShut {NoStop}%
\bibitem [{\citenamefont {Schumm}\ \emph {et~al.}(2005)\citenamefont {Schumm},
  \citenamefont {Hofferberth}, \citenamefont {Andersson}, \citenamefont
  {Wildermuth}, \citenamefont {Groth}, \citenamefont {Bar-Joseph},
  \citenamefont {Schmiedmayer},\ and\ \citenamefont
  {Kr{\"u}ger}}]{schumm2005matter}%
  \BibitemOpen
  \bibfield  {author} {\bibinfo {author} {\bibfnamefont {T.}~\bibnamefont
  {Schumm}}, \bibinfo {author} {\bibfnamefont {S.}~\bibnamefont {Hofferberth}},
  \bibinfo {author} {\bibfnamefont {L.~M.}\ \bibnamefont {Andersson}}, \bibinfo
  {author} {\bibfnamefont {S.}~\bibnamefont {Wildermuth}}, \bibinfo {author}
  {\bibfnamefont {S.}~\bibnamefont {Groth}}, \bibinfo {author} {\bibfnamefont
  {I.}~\bibnamefont {Bar-Joseph}}, \bibinfo {author} {\bibfnamefont
  {J.}~\bibnamefont {Schmiedmayer}}, \ and\ \bibinfo {author} {\bibfnamefont
  {P.}~\bibnamefont {Kr{\"u}ger}},\ }\href@noop {} {\bibfield  {journal}
  {\bibinfo  {journal} {Nature physics}\ }\textbf {\bibinfo {volume} {1}},\
  \bibinfo {pages} {57} (\bibinfo {year} {2005})}\BibitemShut {NoStop}%
\bibitem [{\citenamefont {Zibold}\ \emph {et~al.}(2010)\citenamefont {Zibold},
  \citenamefont {Nicklas}, \citenamefont {Gross},\ and\ \citenamefont
  {Oberthaler}}]{Zibold}%
  \BibitemOpen
  \bibfield  {author} {\bibinfo {author} {\bibfnamefont {T.}~\bibnamefont
  {Zibold}}, \bibinfo {author} {\bibfnamefont {E.}~\bibnamefont {Nicklas}},
  \bibinfo {author} {\bibfnamefont {C.}~\bibnamefont {Gross}}, \ and\ \bibinfo
  {author} {\bibfnamefont {M.~K.}\ \bibnamefont {Oberthaler}},\ }\href
  {\doibase 10.1103/PhysRevLett.105.204101} {\bibfield  {journal} {\bibinfo
  {journal} {Phys. Rev. Lett.}\ }\textbf {\bibinfo {volume} {105}},\ \bibinfo
  {pages} {204101} (\bibinfo {year} {2010})}\BibitemShut {NoStop}%
\bibitem [{\citenamefont {Maussang}\ \emph {et~al.}(2010)\citenamefont
  {Maussang}, \citenamefont {Marti}, \citenamefont {Schneider}, \citenamefont
  {Treutlein}, \citenamefont {Li}, \citenamefont {Sinatra}, \citenamefont
  {Long}, \citenamefont {Est\`eve},\ and\ \citenamefont {Reichel}}]{Maussang}%
  \BibitemOpen
  \bibfield  {author} {\bibinfo {author} {\bibfnamefont {K.}~\bibnamefont
  {Maussang}}, \bibinfo {author} {\bibfnamefont {G.~E.}\ \bibnamefont {Marti}},
  \bibinfo {author} {\bibfnamefont {T.}~\bibnamefont {Schneider}}, \bibinfo
  {author} {\bibfnamefont {P.}~\bibnamefont {Treutlein}}, \bibinfo {author}
  {\bibfnamefont {Y.}~\bibnamefont {Li}}, \bibinfo {author} {\bibfnamefont
  {A.}~\bibnamefont {Sinatra}}, \bibinfo {author} {\bibfnamefont
  {R.}~\bibnamefont {Long}}, \bibinfo {author} {\bibfnamefont {J.}~\bibnamefont
  {Est\`eve}}, \ and\ \bibinfo {author} {\bibfnamefont {J.}~\bibnamefont
  {Reichel}},\ }\href {\doibase 10.1103/PhysRevLett.105.080403} {\bibfield
  {journal} {\bibinfo  {journal} {Phys. Rev. Lett.}\ }\textbf {\bibinfo
  {volume} {105}},\ \bibinfo {pages} {080403} (\bibinfo {year}
  {2010})}\BibitemShut {NoStop}%
\bibitem [{\citenamefont {Doria}\ \emph {et~al.}(2011)\citenamefont {Doria},
  \citenamefont {Calarco},\ and\ \citenamefont {Montangero}}]{CRAB}%
  \BibitemOpen
  \bibfield  {author} {\bibinfo {author} {\bibfnamefont {P.}~\bibnamefont
  {Doria}}, \bibinfo {author} {\bibfnamefont {T.}~\bibnamefont {Calarco}}, \
  and\ \bibinfo {author} {\bibfnamefont {S.}~\bibnamefont {Montangero}},\
  }\href {\doibase 10.1103/PhysRevLett.106.190501} {\bibfield  {journal}
  {\bibinfo  {journal} {Phys. Rev. Lett.}\ }\textbf {\bibinfo {volume} {106}},\
  \bibinfo {pages} {190501} (\bibinfo {year} {2011})}\BibitemShut {NoStop}%
\bibitem [{\citenamefont {Allahverdyan}\ and\ \citenamefont
  {Nieuwenhuizen}(2005)}]{Allahverdyan2005}%
  \BibitemOpen
  \bibfield  {author} {\bibinfo {author} {\bibfnamefont {A.~E.}\ \bibnamefont
  {Allahverdyan}}\ and\ \bibinfo {author} {\bibfnamefont {T.~M.}\ \bibnamefont
  {Nieuwenhuizen}},\ }\href {\doibase 10.1103/PhysRevE.71.046107} {\bibfield
  {journal} {\bibinfo  {journal} {Phys. Rev. E}\ }\textbf {\bibinfo {volume}
  {71}},\ \bibinfo {pages} {046107} (\bibinfo {year} {2005})}\BibitemShut
  {NoStop}%
\bibitem [{\citenamefont {Lipkin}\ \emph {et~al.}(1965)\citenamefont {Lipkin},
  \citenamefont {Meshkov},\ and\ \citenamefont {Glick}}]{lipkin1965validity}%
  \BibitemOpen
  \bibfield  {author} {\bibinfo {author} {\bibfnamefont {H.~J.}\ \bibnamefont
  {Lipkin}}, \bibinfo {author} {\bibfnamefont {N.}~\bibnamefont {Meshkov}}, \
  and\ \bibinfo {author} {\bibfnamefont {A.}~\bibnamefont {Glick}},\
  }\href@noop {} {\bibfield  {journal} {\bibinfo  {journal} {Nuclear Physics}\
  }\textbf {\bibinfo {volume} {62}},\ \bibinfo {pages} {188} (\bibinfo {year}
  {1965})}\BibitemShut {NoStop}%
\bibitem [{\citenamefont {Ribeiro}\ \emph {et~al.}(2008)\citenamefont
  {Ribeiro}, \citenamefont {Vidal},\ and\ \citenamefont {Mosseri}}]{Vidal}%
  \BibitemOpen
  \bibfield  {author} {\bibinfo {author} {\bibfnamefont {P.}~\bibnamefont
  {Ribeiro}}, \bibinfo {author} {\bibfnamefont {J.}~\bibnamefont {Vidal}}, \
  and\ \bibinfo {author} {\bibfnamefont {R.}~\bibnamefont {Mosseri}},\ }\href
  {\doibase 10.1103/PhysRevE.78.021106} {\bibfield  {journal} {\bibinfo
  {journal} {Phys. Rev. E}\ }\textbf {\bibinfo {volume} {78}},\ \bibinfo
  {pages} {021106} (\bibinfo {year} {2008})}\BibitemShut {NoStop}%
\bibitem [{\citenamefont {Brennecke}\ \emph {et~al.}(2013)\citenamefont
  {Brennecke}, \citenamefont {Mottl}, \citenamefont {Baumann}, \citenamefont
  {Landig}, \citenamefont {Donner},\ and\ \citenamefont
  {Esslinger}}]{brennecke2013real}%
  \BibitemOpen
  \bibfield  {author} {\bibinfo {author} {\bibfnamefont {F.}~\bibnamefont
  {Brennecke}}, \bibinfo {author} {\bibfnamefont {R.}~\bibnamefont {Mottl}},
  \bibinfo {author} {\bibfnamefont {K.}~\bibnamefont {Baumann}}, \bibinfo
  {author} {\bibfnamefont {R.}~\bibnamefont {Landig}}, \bibinfo {author}
  {\bibfnamefont {T.}~\bibnamefont {Donner}}, \ and\ \bibinfo {author}
  {\bibfnamefont {T.}~\bibnamefont {Esslinger}},\ }\href@noop {} {\bibfield
  {journal} {\bibinfo  {journal} {Proceedings of the National Academy of
  Sciences}\ }\textbf {\bibinfo {volume} {110}},\ \bibinfo {pages} {11763}
  (\bibinfo {year} {2013})}\BibitemShut {NoStop}%
\bibitem [{\citenamefont {Ford}\ \emph {et~al.}(2012)\citenamefont {Ford},
  \citenamefont {Minor},\ and\ \citenamefont {Binnie}}]{Ford}%
  \BibitemOpen
  \bibfield  {author} {\bibinfo {author} {\bibfnamefont {I.~J.}\ \bibnamefont
  {Ford}}, \bibinfo {author} {\bibfnamefont {D.~S.}\ \bibnamefont {Minor}}, \
  and\ \bibinfo {author} {\bibfnamefont {S.~J.}\ \bibnamefont {Binnie}},\
  }\href {http://stacks.iop.org/0143-0807/33/i=6/a=1789} {\bibfield  {journal}
  {\bibinfo  {journal} {European Journal of Physics}\ }\textbf {\bibinfo
  {volume} {33}},\ \bibinfo {pages} {1789} (\bibinfo {year}
  {2012})}\BibitemShut {NoStop}%
\bibitem [{\citenamefont {Deffner}\ and\ \citenamefont
  {Lutz}(2008)}]{DeffnerLutz}%
  \BibitemOpen
  \bibfield  {author} {\bibinfo {author} {\bibfnamefont {S.}~\bibnamefont
  {Deffner}}\ and\ \bibinfo {author} {\bibfnamefont {E.}~\bibnamefont {Lutz}},\
  }\href {\doibase 10.1103/PhysRevE.77.021128} {\bibfield  {journal} {\bibinfo
  {journal} {Phys. Rev. E}\ }\textbf {\bibinfo {volume} {77}},\ \bibinfo
  {pages} {021128} (\bibinfo {year} {2008})}\BibitemShut {NoStop}%
\bibitem [{\citenamefont {Salamon}\ \emph {et~al.}(2009)\citenamefont
  {Salamon}, \citenamefont {Hoffmann}, \citenamefont {Rezek},\ and\
  \citenamefont {Kosloff}}]{Kosloff2009}%
  \BibitemOpen
  \bibfield  {author} {\bibinfo {author} {\bibfnamefont {P.}~\bibnamefont
  {Salamon}}, \bibinfo {author} {\bibfnamefont {K.~H.}\ \bibnamefont
  {Hoffmann}}, \bibinfo {author} {\bibfnamefont {Y.}~\bibnamefont {Rezek}}, \
  and\ \bibinfo {author} {\bibfnamefont {R.}~\bibnamefont {Kosloff}},\ }\href
  {\doibase 10.1039/B816102J} {\bibfield  {journal} {\bibinfo  {journal} {Phys.
  Chem. Chem. Phys.}\ }\textbf {\bibinfo {volume} {11}},\ \bibinfo {pages}
  {1027} (\bibinfo {year} {2009})}\BibitemShut {NoStop}%
\bibitem [{\citenamefont {Galve}\ and\ \citenamefont {Lutz}(2009)}]{GalveLutz}%
  \BibitemOpen
  \bibfield  {author} {\bibinfo {author} {\bibfnamefont {F.}~\bibnamefont
  {Galve}}\ and\ \bibinfo {author} {\bibfnamefont {E.}~\bibnamefont {Lutz}},\
  }\href {\doibase 10.1103/PhysRevA.79.032327} {\bibfield  {journal} {\bibinfo
  {journal} {Phys. Rev. A}\ }\textbf {\bibinfo {volume} {79}},\ \bibinfo
  {pages} {032327} (\bibinfo {year} {2009})}\BibitemShut {NoStop}%
\bibitem [{\citenamefont {Bonan\c{c}a}\ and\ \citenamefont
  {Deffner}(2014)}]{BonancaDeffner}%
  \BibitemOpen
  \bibfield  {author} {\bibinfo {author} {\bibfnamefont {M.~V.~S.}\
  \bibnamefont {Bonan\c{c}a}}\ and\ \bibinfo {author} {\bibfnamefont
  {S.}~\bibnamefont {Deffner}},\ }\href
  {http://scitation.aip.org/content/aip/journal/jcp/140/24/10.1063/1.4885277}
  {\bibfield  {journal} {\bibinfo  {journal} {The Journal of Chemical Physics}\
  }\textbf {\bibinfo {volume} {140}},\ \bibinfo {eid} {244119} (\bibinfo {year}
  {2014})}\BibitemShut {NoStop}%
\bibitem [{\citenamefont {Krotov}(1995)}]{krotov1995global}%
  \BibitemOpen
  \bibfield  {author} {\bibinfo {author} {\bibfnamefont {V.}~\bibnamefont
  {Krotov}},\ }\href@noop {} {\emph {\bibinfo {title} {Global methods in
  optimal control theory}}},\ Vol.\ \bibinfo {volume} {195}\ (\bibinfo
  {publisher} {CRC Press},\ \bibinfo {year} {1995})\BibitemShut {NoStop}%
\bibitem [{\citenamefont {Moore}\ and\ \citenamefont {Rabitz}(2011)}]{Rabitz}%
  \BibitemOpen
  \bibfield  {author} {\bibinfo {author} {\bibfnamefont {K.~W.}\ \bibnamefont
  {Moore}}\ and\ \bibinfo {author} {\bibfnamefont {H.}~\bibnamefont {Rabitz}},\
  }\href {\doibase 10.1103/PhysRevA.84.012109} {\bibfield  {journal} {\bibinfo
  {journal} {Phys. Rev. A}\ }\textbf {\bibinfo {volume} {84}},\ \bibinfo
  {pages} {012109} (\bibinfo {year} {2011})}\BibitemShut {NoStop}%
\bibitem [{\citenamefont {van Frank}\ \emph {et~al.}(2014)\citenamefont {van
  Frank}, \citenamefont {Negretti}, \citenamefont {Berrada}, \citenamefont
  {B{\"u}cker}, \citenamefont {Montangero}, \citenamefont {Schaff},
  \citenamefont {Schumm}, \citenamefont {Calarco},\ and\ \citenamefont
  {Schmiedmayer}}]{Schmiedmayer2014}%
  \BibitemOpen
  \bibfield  {author} {\bibinfo {author} {\bibfnamefont {S.}~\bibnamefont {van
  Frank}}, \bibinfo {author} {\bibfnamefont {A.}~\bibnamefont {Negretti}},
  \bibinfo {author} {\bibfnamefont {T.}~\bibnamefont {Berrada}}, \bibinfo
  {author} {\bibfnamefont {R.}~\bibnamefont {B{\"u}cker}}, \bibinfo {author}
  {\bibfnamefont {S.}~\bibnamefont {Montangero}}, \bibinfo {author}
  {\bibfnamefont {J.~F.}\ \bibnamefont {Schaff}}, \bibinfo {author}
  {\bibfnamefont {T.}~\bibnamefont {Schumm}}, \bibinfo {author} {\bibfnamefont
  {T.}~\bibnamefont {Calarco}}, \ and\ \bibinfo {author} {\bibfnamefont
  {J.}~\bibnamefont {Schmiedmayer}},\ }\href
  {http://dx.doi.org/10.1038/ncomms5009} {\bibfield  {journal} {\bibinfo
  {journal} {Nat Commun}\ }\textbf {\bibinfo {volume} {5}} (\bibinfo {year}
  {2014})}\BibitemShut {NoStop}%
\bibitem [{\citenamefont {Auccaise}\ \emph {et~al.}(2015)\citenamefont
  {Auccaise}, \citenamefont {Araujo-Ferreira}, \citenamefont {Sarthour},
  \citenamefont {Oliveira}, \citenamefont {Bonagamba},\ and\ \citenamefont
  {Roditi}}]{Auccaise}%
  \BibitemOpen
  \bibfield  {author} {\bibinfo {author} {\bibfnamefont {R.}~\bibnamefont
  {Auccaise}}, \bibinfo {author} {\bibfnamefont {A.~G.}\ \bibnamefont
  {Araujo-Ferreira}}, \bibinfo {author} {\bibfnamefont {R.~S.}\ \bibnamefont
  {Sarthour}}, \bibinfo {author} {\bibfnamefont {I.~S.}\ \bibnamefont
  {Oliveira}}, \bibinfo {author} {\bibfnamefont {T.~J.}\ \bibnamefont
  {Bonagamba}}, \ and\ \bibinfo {author} {\bibfnamefont {I.}~\bibnamefont
  {Roditi}},\ }\href {\doibase 10.1103/PhysRevLett.114.043604} {\bibfield
  {journal} {\bibinfo  {journal} {Phys. Rev. Lett.}\ }\textbf {\bibinfo
  {volume} {114}},\ \bibinfo {pages} {043604} (\bibinfo {year}
  {2015})}\BibitemShut {NoStop}%
\bibitem [{\citenamefont {Araujo-Ferreira}\ \emph {et~al.}(2013)\citenamefont
  {Araujo-Ferreira}, \citenamefont {Auccaise}, \citenamefont {Sarthour},
  \citenamefont {Oliveira}, \citenamefont {Bonagamba},\ and\ \citenamefont
  {Roditi}}]{Auccaise2}%
  \BibitemOpen
  \bibfield  {author} {\bibinfo {author} {\bibfnamefont {A.~G.}\ \bibnamefont
  {Araujo-Ferreira}}, \bibinfo {author} {\bibfnamefont {R.}~\bibnamefont
  {Auccaise}}, \bibinfo {author} {\bibfnamefont {R.~S.}\ \bibnamefont
  {Sarthour}}, \bibinfo {author} {\bibfnamefont {I.~S.}\ \bibnamefont
  {Oliveira}}, \bibinfo {author} {\bibfnamefont {T.~J.}\ \bibnamefont
  {Bonagamba}}, \ and\ \bibinfo {author} {\bibfnamefont {I.}~\bibnamefont
  {Roditi}},\ }\href {\doibase 10.1103/PhysRevA.87.053605} {\bibfield
  {journal} {\bibinfo  {journal} {Phys. Rev. A}\ }\textbf {\bibinfo {volume}
  {87}},\ \bibinfo {pages} {053605} (\bibinfo {year} {2013})}\BibitemShut
  {NoStop}%
\bibitem [{\citenamefont {Husimi}(1953)}]{Husimi01041953}%
  \BibitemOpen
  \bibfield  {author} {\bibinfo {author} {\bibfnamefont {K.}~\bibnamefont
  {Husimi}},\ }\href {\doibase 10.1143/ptp/9.4.381} {\bibfield  {journal}
  {\bibinfo  {journal} {Progress of Theoretical Physics}\ }\textbf {\bibinfo
  {volume} {9}},\ \bibinfo {pages} {381} (\bibinfo {year} {1953})}\BibitemShut
  {NoStop}%
\end{thebibliography}%


%


\end{document}